\documentstyle[12pt,preprint,aps]{revtex}

\newcommand{\csi}{{\rm cos}(\Theta^\pm_p)}
\newcommand{\csii}{{\rm cos}^2(\Theta^\pm_p)}
\newcommand{\gvp}{\tilde{g}_{\rm v}\left(p\right)}
\newcommand{\fvp}{\tilde{f}_{\rm v}\left(p\right)}
\newcommand{\gip}{\tilde{g}^{\left(1\right)}\left(p\right)}
\newcommand{\fip}{\tilde{f}^{\left(1\right)}\left(p\right)}
\newcommand{\giip}{\tilde{g}^{\left(2\right)}\left(p\right)}
\newcommand{\fiip}{\tilde{f}^{\left(2\right)}\left(p\right)}
\newcommand{\fup}{\tilde{f}_\mu^{\left(i\right)}\left(p\right)}
\newcommand{\fupq}{\tilde{f}^{\left(i\right)}\left(p\right)}
\newcommand{\gup}{\tilde{g}_\mu^{\left(i\right)}\left(p\right)}
\newcommand{\gupq}{\tilde{g}^{\left(i\right)}\left(p\right)}
\newcommand{\guip}{\tilde{g}_{\mu}^{\left(1\right)}\left(p\right)}
\newcommand{\fuip}{\tilde{f}_{\mu}^{\left(1\right)}\left(p\right)}
\newcommand{\guiip}{\tilde{g}_{\mu}^{\left(2\right)}\left(p\right)}
\newcommand{\fuiip}{\tilde{f}_{\mu}^{\left(2\right)}\left(p\right)}

\newcommand{\gvr}{g_{\rm v}\left(r\right)}
\newcommand{\fvr}{f_{\rm v}\left(r\right)}
\newcommand{\guiir}{g_{\mu}^{\left(2\right)}\left(r\right)}
\newcommand{\fuiir}{f_{\mu}^{\left(2\right)}\left(r\right)}
\newcommand{\be}{\begin{eqnarray}}
\newcommand{\ba}{\begin{array}}
\newcommand{\ea}{\end{array}}
\newcommand{\ee}{\end{eqnarray}}

\newcommand{\dslash}{\partial \hskip -0.5em /}

\newcommand{\Tr}{{\rm Tr}}

\newcommand{\La}{{\cal L}}
\newcommand{\A}{{\cal A}}

\newcommand{\zr}[1]{\mbox{\hspace*{#1em}}}
\newcommand{\ID}{\mbox{{\sf 1}\zr{-0.16}\rule{0.04em}{1.55ex}\zr{0.1}}}
 
\begin{document}
\preprint{\vbox{\hbox{January 1997}}}
\vskip 1.0truecm
\title{Polarized Nucleon Structure Functions
within a Chiral Soliton Model}
\vskip 1.0cm
\author{H.\ Weigel, L.\ Gamberg\footnote{
Present address:
\parbox[t]{11cm}{Department of Physics and Astronomy,
University of Oklahoma, 440 West Brooks, Norman, Ok 73019, USA}},
and H.\ Reinhardt}
\address{Institute for Theoretical Physics
T\"ubingen University\\
Auf der Morgenstelle 14, D-72076 T\"ubingen, Germany}
\maketitle
\begin{abstract}
We study polarized--spin structure functions of the nucleon
within the bosonized 
Nambu--Jona--Lasinio model where the nucleon emerges as a 
chiral soliton. We present the electromagnetic polarized structure 
functions, $g_{1}\left(x\right)$ and  $g_{2}\left(x\right)$ for 
$ep$ scattering and discuss various sum rules in the valence quark 
approximation. This approximation is justified because in this model 
axial properties of the nucleon are dominated by their valence quark 
contributions.  We find that these structure functions are well 
localized in the interval $0\le x \le1$. We compare the model 
predictions on the polarized structure functions with data from the
E143 experiment by evolving them from the scale characteristic of the 
NJL-model to the scale of the data. Additionally a comparison is made
with parameterized data at a momentum scale commensurate with the 
model calculation.
\end{abstract}

\vskip 0.5cm
\leftline{\it PACS: 12.39.Fe, 12.39.Ki.}
\vskip 0.5cm

\newpage

\normalsize\baselineskip=20pt
\section{Introduction}
\bigskip

Over the past decade, beginning with the measurement of nucleon
spin--polarized structure function, $g_{1}(x,Q^2)$ by the 
EMC \cite{emc88} at CERN and most recently with the 
spin--structure function $g_2(x,Q^2)$ in the E143 
experiment \cite{slac96} at SLAC, a wealth of information has been 
gathered on the spin--polarized structure functions of the nucleon 
and their corresponding sum rules (see in addition
\cite{smc93}, \cite{smc94a}, \cite{smc94b}, \cite{slac93}, 
\cite{slac95a}, \cite{slac95b}).
Initially the analysis of these experiments cast doubt 
on the non--relativistic quark model \cite{kok79} 
interpretations regarding the spin content of the proton. 
By now it is firmly established 
that the quark helicity of the nucleon is much smaller than the 
predictions of that model, however, many questions remain
to be addressed concerning the spin structure.
As a result there have been numerous investigations within 
models for the nucleon in an effort to determine the manner in which
the nucleon spin is distributed among its constituents.
One option is to study 
the axial current matrix elements of the 
nucleon such as
$\langle N |{\cal A}_{\mu}^{i}|N\rangle = 2\Delta q_{i}S_{\mu}$,
which, for example, provide information on 
the nucleon axial singlet charge
\be
g_{A}^{0}&=&\langle N |{\cal A}_{3}^{0}|N\rangle\ 
= \left( \Delta u + \Delta d + \Delta s\right)=
\Gamma_{1}^{p} (Q^2) + \Gamma_{1}^{n} (Q^2) \ .
\ee
Here $\Delta q$ are the axial charges of the quark constituents and  
$\Gamma_{1}^{N} (Q^2)=\int_{0}^{1} dx g_{1}^{N}(x,Q^2)$ is the
first moment of the longitudinal nucleon spin structure function,
$g_1^N\left(x,Q^{2}\right)$.
Of course, it is more illuminating to 
directly compute the longitudinal and transverse 
nucleon  spin--structure functions, $g_{1}\left(x,Q^2\right)$
and $g_{T}(x,Q^2)=g_{1}(x,Q^2)+g_{2}(x,Q^2)$, respectively as
functions of the Bjorken variable $x$.
We will calculate these structure functions within the
Nambu--Jona--Lasinio (NJL) \cite{Na61} chiral soliton model \cite{Re88}.

Chiral soliton models are unique both in being 
the first effective models of hadronic physics to shed light 
on the so called  ``proton--spin crisis" by predicting a singlet 
combination in accord with the data \cite{br88}, and in predicting a 
non--trivial strange quark content to the axial vector current of the 
nucleon \cite{br88}, \cite{pa89}, \cite{jon90}, \cite{blo93}; 
about $10-30\%$ of the down quarks 
(see \cite{wei96a} and \cite{ell96} for reviews).  
However, while the leading moments of these
structure functions have been calculated within chiral  soliton 
models, from the Skyrme model \cite{Sk61}, \cite{Ad83} and its various 
vector--meson extensions, to
models containing explicit quark degrees of freedom such as the 
(NJL) model \cite{Na61}, 
the nucleon spin--structure functions 
themselves have not been investigated in these models.  
Soliton model calculations of structure functions
were, however, performed in 
Friedberg-Lee \cite{frie77} and color-dielectric \cite{nil82} models. 
In addition, structure functions have extensively been studied
within the framework of effective quark models such as the 
bag--model \cite{Ch74}, and the Center of Mass  
bag model \cite{So94}. 
These models are confining by construction but they
neither contain non--perturbative pseudoscalar fields nor are they
chirally symmetric\footnote{In the
cloudy bag model the contribution of the pions to structure 
functions has at most been treated perturbatively
\cite{Sa88}, \cite{Sc92}.}.
To this date it is fair to say that 
many of the successes of low--energy effective models rely on
the incorporation of chiral symmetry and its spontaneous 
symmetry breaking (see for e.g. \cite{Al96}).
In this article we therefore present our 
calculation of the polarized spin structure functions in the NJL
chiral soliton model \cite{Re89}, \cite{Al96}.  
Since in particular the static axial properties of the nucleon are
dominated by the valence quark contribution in this model
it is legitimate to focus on the valence quarks in this model.

At the outset it is important to note that a major difference 
between the chiral soliton models and models previously employed 
to calculate structure functions is the form of the nucleon 
wave--function. In the latter the nucleon wave--function is a 
product of Dirac spinors while in the former the nucleon appears as 
a collectively excited (topologically) non--trivial meson 
configuration. 

As in the original bag model study \cite{Ja75} 
of structure functions for 
localized field configurations, the structure functions are most
easily accessible when the current operator is at most quadratic in 
the fundamental fields and the propagation of the interpolating 
field can be regarded as free. 
Although the latter approximation is 
well justified in the Bjorken limit the former condition is 
difficult to satisfy in soliton models where mesons 
are fundamental fields ({\it e.g.} the Skyrme 
model \cite{Sk61}, \cite{Ad83}, 
the chiral quark model of ref. \cite{Bi85} or the chiral bag model 
\cite{Br79}). 
Such model Lagrangians typically possess all orders of the fundamental
pion field.  In that case the current operator is not confined to 
quadratic order and the calculation of the hadronic tensor 
(see eq. (\ref{deften}) below) requires drastic approximations.
In this respect the chirally invariant NJL model 
is preferred because it is entirely defined in terms of quark 
degrees of freedom and formally the current
possesses the structure as in a non--interacting 
model.  This makes the evaluation of the hadronic tensor
feasible.
Nevertheless after bosonization 
the hadronic currents 
are uniquely defined 
functionals of the solitonic meson fields.

The paper is organized as follows: In section 2 we give a brief
discussion of the standard operator product expansion (OPE) analysis 
to establish the connection between the effective models for the 
baryons at low energies and the quark--parton model description. 
In section 3 we briefly review the NJL chiral soliton.
In section 4 we extract the polarized structure 
functions from the hadronic tensor, eq. (\ref{had})
exploiting the ``valence quark approximation".
Section 5 displays the results of the spin--polarized 
structure functions calculated in the  NJL chiral soliton model 
within this approximation and compare this result with a recent 
low--renormalization point parametrization \cite{Gl95}. 
In section 6 we use Jaffe's prescription \cite{Ja80} to impose 
proper support for the structure 
function within the interval $x\in \left[0,1\right]$. 
Subsequently the structure functions 
are evolved \cite{Al73}, \cite{Al94}, \cite{Ali91} 
from the scale characterizing the NJL--model to the
scale associated with the experimental data.  Section 7 serves to 
summarize these studies and to propose further explorations. 
In appendix A we list explicit analytic expressions for the isoscalar 
and isovector polarized structure functions.
Appendix B summarizes details on the evolution of the twist--3
structure function, ${\overline{g}}_2\left(x,Q^2\right)$.

\bigskip
\section{DIS and the Chiral Soliton}
\bigskip

It has been a long standing effort to establish the connection 
between the chiral soliton picture of the baryon, which essentially 
views baryons as mesonic lumps and the quark parton model which 
regards baryons as composites of almost non--interacting, point--like 
quarks. While the former has been quite successful in describing 
static properties of the nucleon, the latter, being firmly 
established within the context of deep inelastic scattering (DIS), 
has been employed  extensively to calculate the short distance or
perturbative processes within QCD.
In fact this connection can be made through the OPE.

The discussion begins with  the hadronic tensor for electron--nucleon 
scattering, 
\be
W_{\mu\nu}(q)=\frac{1}{4\pi}\int d^4 \xi \ 
{\rm e}^{iq\cdot\xi}
\langle N |\left[J_\mu(\xi),J^{\dag}_\nu(0)\right]|N\rangle\ ,
\label{deften}
\ee
where $J_\mu={\bar q}(\xi)\gamma_\mu {\cal Q} q(\xi)$ is the
electromagnetic 
current, ${\cal Q}=\left(\frac{2}{3},\frac{-1}{3}\right)$ is the (two
flavor) quark charge matrix and $|N\rangle$ refers to the 
nucleon state.  In the DIS regime the OPE enables one to 
express the product of these 
currents in terms of the forward Compton scattering 
amplitude $T_{\mu\nu}(q)$ of a virtual photon 
from a nucleon 
\be
T_{\mu\nu}(q)=i\int d^4 \xi \ 
{\rm e}^{iq\cdot\xi}
\langle N |T\left(J_\mu(\xi)J^{\dag}_\nu(0)\right)|N\rangle\ ,
\label{im}
\ee
by an expansion on the light cone  $\left(\xi^2 \rightarrow 0\right)$
using a set of renormalized local 
operators \cite{muta87}, \cite{rob90}.  In the Bjorken 
limit the influence of these operators is determined 
by the twist, $\tau$ or the light cone singularity of their coefficient 
functions.  Effectively this becomes a power series in the inverse
of the Bjorken variable $x=-q^{2}/2P\cdot q$, with $P_\mu$ being the
nucleon momentum:
\be
T_{\mu\nu}(q)\ =\sum_{n,i,\tau}
\left(\frac{1}{x}\right)^{n}\ e_{\mu\nu}^{i}\left(q,P,S\right)\ 
C^{n}_{\tau,i}(Q^2/\mu^2,\alpha_s(\mu^2)){\cal O}^{n}_{\tau,i}(\mu^2)
(\frac{1}{Q^2})^{\frac{\tau}{2}\ - 1}\ .
\label{series}
\ee
Here the  index $i$ runs over all scalar matrix 
elements, ${\cal O}_{\tau,i}^{n}(\mu^2)$, with the 
same Lorentz structure (characterized by
the tensor, $e_{\mu\nu}^{i}$).  Furthermore,
$S^{\mu}$ is the spin 
of the nucleon, 
$\left(S^2=-1\ , S\cdot P\ =0\right)$ and $Q^2=-q^2 > 0$.  
As is evident, higher twist contributions
are suppressed by powers of $1/{Q^2}$.
The coefficient functions,
$C^{n}_{\tau,i}(Q^2/\mu^2,\alpha_s(\mu^2))$
are target independent and in principle include all 
QCD radiative corrections. Their $Q^2$
variation is determined from the solution of the renormalization 
group equations and logarithmically diminishes at large $Q^2$.
On the other hand the reduced--matrix elements, 
${\cal O}_{\tau,i}^{n}(\mu^2)$,
depend only on the renormalization 
scale $\mu^2$ and reflect the non--perturbative properties
of the nucleon \cite{ans95}.

The optical theorem states that the hadronic tensor is
given in terms of the imaginary part of the virtual Compton scattering
amplitude, $W_{\mu\nu}=\frac{1}{2\pi}{\rm Im}\ T_{\mu\nu}$. 
From the analytic properties of $T_{\mu\nu}(q)$,
together with eq. (\ref{series})  an infinite set of sum rules 
result for the form factors, ${\cal W}_{i}\left(x,Q^2\right)$,
which are defined via the Lorentz covariant decomposition
$W_{\mu\nu}(q)=e_{\mu\nu}^{i}{\cal W}_i\left(x,Q^2\right)$.
These sum rules read
\be 
\int^{1}_{0}dx\ x^{n-1}\ {\cal W}_{i}\left(y,Q^2\right)&=& 
\sum _{\tau}\ 
C^{n}_{\tau,i}\left(Q^2/\mu^2,\alpha_s(\mu^2)\right)
{\cal O}_{\tau,i}^{n}(\mu^2)
(\frac{1}{Q^2})^{\frac{\tau}{2}\ - 1}\ .
\label{ope}
\ee
In the 
{\em impulse approximation}
(i.e. neglecting radiative 
corrections) \cite{Ja90,Ji90,Ja91}
one can directly sum the OPE gaining direct access 
to the structure functions in terms of the reduced matrix elements
${\cal O}_{\tau,i}^{n}(\mu^2)$.

When calculating the renormalization--scale dependent 
matrix elements, ${\cal O}_{\tau,i}^{n}(\mu^2)$ 
within QCD, $\mu^2$ is an arbitrary parameter adjusted to ensure 
rapid convergence of the perturbation series.
However, given the difficulties of obtaining a 
satisfactory description of the nucleon
as a bound--state in the $Q^2$ regime of DIS processes 
it is customary to calculate these 
matrix elements in models at a 
low scale $\mu^2$ and subsequently evolve these results 
to the relevant DIS momentum region of the data
employing, for example, the 
Altarelli--Parisi evolution \cite{Al73}, \cite{Al94}.
In this context, the scale, $\mu^2 \sim \Lambda_{QCD}^{2}$,
characterizes the non--perturbative regime 
where it is possible to formulate a nucleon 
wave--function from which structure functions
are computed.

Here we will utilize the NJL chiral--soliton  model to 
calculate the spin--polarized nucleon structure functions
at the scale, $\mu^2$, subsequently evolving 
the structure functions according to the Altarelli--Parisi scheme.
This establishes the connection between chiral soliton and the
parton models. In addition we compare the structure functions 
calculated in the NJL model to a parameterization of spin structure 
function \cite{Gl95}  at a scale commensurate with our model.

\bigskip
\section{The Nucleon State in the NJL Model}
\bigskip

The Lagrangian of the NJL model reads
\be
\La = \bar q (i\dslash -  m^0 ) q +
      2G_{\rm NJL} \sum _{i=0}^{3}
\left( (\bar q \frac {\tau^i}{2} q )^2
      +(\bar q \frac {\tau^i}{2} i\gamma _5 q )^2 \right) .
\label{NJL}
\ee
Here $q$, $\hat m^0$ and $G_{\rm NJL}$ denote the quark field, the 
current quark mass and a dimensionful coupling constant, respectively.
When integration out the gluon fields from QCD a current--current 
interaction remains, which is meditated by the gluon propagator. 
Replacing this gluon propagator by a local contact interaction and 
performing the appropriate Fierz--transformations yields the 
Lagrangian (\ref{NJL}) in leading order of $1/N_c$ \cite{Re90}, 
where $N_c$ refers to the number of color degrees of freedom. It is 
hence apparent that the interaction term in eq. (\ref{NJL}) is a 
remnant of the gluon fields. Hence gluonic effects are included 
in the model described by the Lagrangian (\ref{NJL}).

Application of functional bosonization techniques \cite{Eb86} to the 
Lagrangian (\ref{NJL}) yields the mesonic action
\be
\A&=&\Tr_\Lambda\log(iD)+\frac{1}{4G_{\rm NJL}}
\int d^4x\ {\rm tr}
\left(m^0\left(M+M^{\dag}\right)-MM^{\dag}\right)\ , 
\label{bosact} \\
D&=&i\dslash-\left(M+M^{\dag}\right)
-\gamma_5\left(M-M^{\dag}\right)\ .
\label{dirac}
\ee
The composite scalar ($S$) and pseudoscalar ($P$) meson fields 
are contained in $M=S+iP$ and appear as quark--antiquark bound 
states.  The NJL model embodies the approximate chiral symmetry of QCD
and has to be understood as an effective (non--renormalizable) theory
of the low--energy quark flavor dynamics.
For regularization, which is indicated by the cut--off 
$\Lambda$, we will adopt the proper--time scheme \cite{Sch51}.
The free parameters of the model are the current quark mass $m^0$, 
the coupling constant $G_{\rm NJL}$ and the cut--off $\Lambda$.
Upon expanding $\A$ to quadratic order in $M$ these parameters are 
related to the pion mass, $m_\pi=135{\rm MeV}$ and pion decay constant,
$f_\pi=93{\rm MeV}$. This leaves one undetermined parameter which we 
choose to be the vacuum expectation value $m=\langle M\rangle$. For 
apparent reasons $m$ is called the constituent quark mass. It is 
related to $m^0$, $G_{\rm NJL}$ and $\Lambda$ via the gap--equation, 
{\it i.e.} the equation of motion for the scalar field $S$\cite{Eb86}. The 
occurrence of this vacuum expectation value reflects the spontaneous 
breaking of chiral symmetry and causes the pseudoscalar fields to
emerge as (would--be) Goldstone bosons.

As the NJL model soliton has exhaustively been discussed in 
recent review articles \cite{Al96}, \cite{Gok96} 
we only present those features, 
which are relevant for the computation of the structure functions
in the valence quark approximation.

The chiral soliton is given by the hedgehog configuration 
of the meson fields
\be
M_{\rm H}(\mbox{\boldmath $x$})=m\ {\rm exp}
\left(i\mbox{\boldmath $\tau$}\cdot{\hat{\mbox{\boldmath $x$}}}
\Theta(r)\right)\ .
\label{hedgehog}
\ee
In order to compute the functional trace in eq. (\ref{bosact}) for this 
static configuration we express the 
Dirac operator (\ref{dirac}) as, $D=i\gamma_0(\partial_t-h)$ 
where
\be
h=\mbox{\boldmath $\alpha$}\cdot\mbox{\boldmath $p$}+m\ 
{\rm exp}\left(i\gamma_5\mbox{\boldmath $\tau$}
\cdot{\hat{\mbox{\boldmath $x$}}}\Theta(r)\right)\ 
\label{hamil}
\ee
is the corresponding Dirac Hamiltonian. We denote the eigenvalues 
and eigenfunctions of $h$ by $\epsilon_\mu$ and $\Psi_\mu$, 
respectively. Explicit expressions for these wave--functions are 
displayed in appendix A. In the proper time regularization scheme 
the energy functional of the NJL model is found to be \cite{Re89,Al96}, 
\be
E[\Theta]=
\frac{N_C}{2}\epsilon_{\rm v}
\left(1+{\rm sgn}(\epsilon_{\rm v})\right)
&+&\frac{N_C}{2}\int^\infty_{1/\Lambda^2}
\frac{ds}{\sqrt{4\pi s^3}}\sum_\nu{\rm exp}
\left(-s\epsilon_\nu^2\right)
\nonumber \\* && \hspace{1.5cm}
+\ m_\pi^2 f_\pi^2\int d^3r  \left(1-{\rm cos}\Theta(r)\right) ,
\label{efunct}
\ee
with $N_C=3$ being the number of color degrees of freedom. 
The subscript ``${\rm v}$" denotes the valence quark level. This state 
is the distinct level bound in the soliton background, {\it i.e.}
$-m<\epsilon_{\rm v}<m$. The chiral angle, $\Theta(r)$, is
obtained by self--consistently extremizing $E[\Theta]$ \cite{Re88}.

States possessing good spin and isospin quantum numbers are 
generated by rotating the hedgehog field
\cite{Ad83}
\be
M(\mbox{\boldmath $x$},t)=
A(t)M_{\rm H}(\mbox{\boldmath $x$})A^{\dag}(t)\ ,
\label{collrot}
\ee
which introduces the collective coordinates $A(t)\in SU(2)$. The 
action functional is expanded \cite{Re89} in the angular velocities 
\be 
2A^{\dag}(t)\dot A(t)=
i\mbox{\boldmath $\tau$}\cdot\mbox{\boldmath $\Omega$} \ .
\label{angvel}
\ee
In particular the valence quark wave--function receives a first 
order perturbation
\be
\Psi_{\rm v}(\mbox{\boldmath $x$},t)=
{\rm e}^{-i\epsilon_{\rm v}t}A(t)
\left\{\Psi_{\rm v}(\mbox{\boldmath $x$})
+\frac{1}{2}\sum_{\mu\ne{\rm v}}
\Psi_\mu(\mbox{\boldmath $x$})
\frac{\langle \mu |\mbox{\boldmath $\tau$}\cdot
\mbox{\boldmath $\Omega$}|{\rm v}\rangle}
{\epsilon_{\rm v}-\epsilon_\mu}\right\}=:
{\rm e}^{-i\epsilon_{\rm v}t}A(t)
\psi_{\rm v}(\mbox{\boldmath $x$}).
\label{valrot}
\ee
Here $\psi_{\rm v}(\mbox{\boldmath $x$})$ refers to the spatial part 
of the body--fixed valence quark wave--function with the rotational 
corrections included. Nucleon states $|N\rangle$ are obtained 
by canonical quantization of the collective coordinates, $A(t)$. By 
construction these states live in the Hilbert space of a rigid rotator. 
The eigenfunctions are Wigner $D$--functions
\be
\langle A|N\rangle=\frac{1}{2\pi}
D^{1/2}_{I_3,-J_3}(A)\ ,
\label{nwfct}
\ee
with $I_3$ and $J_3$ being respectively the isospin and spin 
projection quantum numbers of the nucleon.

\bigskip
\section{Polarized Structure Functions in the NJL model}
\bigskip

The starting point for computing nucleon structure functions
is the hadronic tensor, eq. (\ref{deften}).  The polarized structure 
functions are extracted from its antisymmetric 
piece, $W^{(A)}_{\mu\nu}=(W_{\mu\nu}-W_{\nu\mu})/2i$.
Lorentz invariance implies that the
antisymmetric portion, characterizing polarized 
lepton--nucleon scattering, can be decomposed into 
the polarized structure functions, 
$g_1(x,Q^2)$ and $g_2(x,Q^2)$,
\be
W^{(A)}_{\mu\nu}(q)= 
i\epsilon_{\mu\nu\lambda\sigma}\frac{q^{\lambda}M_N}{P\cdot q}
\left\{g_1(x,Q^2)S^{\sigma}+
\left(S^{\sigma}-\frac{q\cdot S}{q\cdot p}P^{\sigma}\right)
g_2(x,Q^2)\right\}\ ,
\label{had}
\ee
again, $P_\mu$ refers to the nucleon momentum and $Q^2=-q^2$. 
The tensors multiplying the structure functions 
in eq. (\ref{had})
should be identified with the Lorentz tensors $e_{\mu\nu}^{i}$ 
in (\ref{series}).

Contracting $W^{(A)}_{\mu\nu}$ with the longitudinal 
$\Lambda^{\mu\nu}_{L}$ and transverse 
$\Lambda^{\mu\nu}_{T}$ projection operators \cite{ans95},
\be
\Lambda^{\mu\nu}_{L}&=&\frac{2}{b}\left\{2P\cdot qxS_{\lambda}+
\frac{1}{q\cdot S}\left[(q\cdot S)^{2}-
\left(\frac{P\cdot q}{M}\right)^2\right] q_{\lambda}\right\}\ P_\tau \
\epsilon^{\mu\nu\lambda\tau },
\label{proj1}\\
\Lambda^{\mu\nu}_{T}&=&\frac{2}{b}
\left\{\left[\left(\frac{P\cdot q}{M}\right)^2+2P\cdot \
qx\right]S_\lambda + \left(q\cdot S\right)q_\lambda\right\}\ P_\tau \
\epsilon^{\mu\nu\lambda\tau }
\label{projT}
\ee
and choosing the pertinent polarization,
yields the longitudinal component 
\be
g_L(x,Q^2)=g_1(x,Q^2)\ ,
\ee 
as well as the transverse combination 
\be
g_T(x,Q^2)=g_1(x,Q^2) + g_2(x,Q^2)\ .
\ee
Also, $b=-4M\left\{\left(\frac{P\cdot q}{M}\right)^2 + 2P\cdot{qx}-
\left(q\cdot{S}\right)^2\right\}$.  In the Bjorken limit, which 
corresponds to the kinematical regime 
\be
q_0=|\mbox{\boldmath $q$}| - M_N x
\quad {\rm with}\quad
|\mbox{\boldmath $q$}|\rightarrow \infty \ ,
\label{bjlimit}
\ee
the antisymmetric component of the hadronic tensor 
becomes \cite{Ja75},
\be
W^{(A)}_{\mu\nu}(q)&=&\int \frac{d^4k}{(2\pi)^4} \
\epsilon_{\mu\rho\nu\sigma}\ k^\rho\
{\rm sgn}\left(k_0\right) \ \delta\left(k^2\right)
\int_{-\infty}^{+\infty} dt \ {\rm e}^{i(k_0+q_0)t}
\nonumber \\* && 
\times \int d^3x_1 \int d^3x_2 \
{\rm exp}\left[-i(\mbox{\boldmath $k$}+\mbox{\boldmath $q$})\cdot
(\mbox{\boldmath $x$}_1-\mbox{\boldmath $x$}_2)\right]
\nonumber \\* && 
\times \langle N |\left\{
{\bar \Psi}(\mbox{\boldmath $x$}_1,t){\cal Q}^2\gamma^\sigma\gamma^{5}
\Psi(\mbox{\boldmath $x$}_2,0)+
{\bar \Psi}(\mbox{\boldmath $x$}_2,0){\cal Q}^2\gamma^\sigma\gamma^{5}
\Psi(\mbox{\boldmath $x$}_1,t)\right\}| N \rangle \ ,
\label{stpnt}
\ee
where $\epsilon_{\mu\rho\nu\sigma}\gamma^\sigma \gamma^5$
is the 
antisymmetric combination of $\gamma_\mu\gamma_\rho\gamma_\nu$.
The matrix element between the nucleon states is to be taken in 
the space of the collective coordinates, $A(t)$ (see eqs. 
(\ref{collrot}) and (\ref{nwfct})) as the object in curly brackets 
is an operator in this space. In deriving the expression (\ref{stpnt})  
the {\it free} correlation function for the intermediate quark 
fields has been assumed\footnote{Adopting a dressed correlation will 
cause corrections starting at order twist--4 in QCD \cite{Ja96}.} 
after applying Wick's theorem to the product of quark currents in eq. 
(\ref{deften}). \cite{Ja75}. The use of the {\it free} correlation
function is justified because in the Bjorken limit (\ref{bjlimit}) 
the intermediate quark fields carry very large momenta and are hence 
not sensitive to typical soliton momenta.  This procedure reduces the 
commutator $[J_\mu(\mbox{\boldmath $x$}_1,t),
J^{\dag}_\nu(\mbox{\boldmath $x$}_2,0)]$ of the quark currents in 
the definition (\ref{deften}) to objects which are merely bilinear 
in the quark fields. Consequently, in the Bjorken limit 
(\ref{bjlimit}) the momentum, $k$, of the intermediate quark state 
is highly off--shell and hence is not sensitive to momenta typical for 
the soliton configuration. Therefore, the use of the free correlation 
function is a good approximation in this kinematical regime.
Accordingly, the intermediate quark states are taken to be massless,
{\it cf.} eq. (\ref{stpnt}).

Since the NJL model is originally defined in terms of quark degrees of 
freedom, quark bilinears as in eq. (\ref{stpnt}) can be computed 
from the functional
\be
\hspace{-1cm}
\langle {\bar q}(x){\cal Q}^{2} q(y) \rangle&=&
\int D{\bar q} Dq \ {\bar q}(x){\cal Q}^{2} q(y)\ 
{\rm exp}\left(i \int d^4x^\prime\ \La\right)
\nonumber \\*  
&=&\frac{\delta}{i\delta\alpha(x,y)}\int D{\bar q} Dq \
{\rm exp}\left(i\int d^4x^\prime d^4y^\prime
\left[\delta^4(x^\prime - y^\prime )\La \right. \right.
\nonumber \\* && \hspace{5cm}
\left. \left.
+\  \alpha(x^\prime,y^\prime){\bar q}(x^\prime){\cal Q}^{2}
q(y^\prime)\right] \right)\Big|_{\alpha(x,y)=0}\ .
\label{gendef}
\ee
The introduction of the bilocal source $\alpha(x,y)$ facilitates 
the  functional bosonization after which eq. (\ref{gendef}) 
takes the form
\be
\frac{\delta}{\delta\alpha(x,y)}{\rm Tr}_{\Lambda}{\rm log}
\left(\delta^4(x-y)D+\alpha(x,y){\cal Q}^{2})\right)
\Big|_{\alpha(x,y)=0}\ \ .
\label{gendef1}
\ee
The operator $D$ is defined in eq. (\ref{dirac}). 
The correlation $\langle {\bar q}(x){\cal Q}^2 q(y) \rangle$ depends on 
the angle between $\mbox{\boldmath $x$}$ and $\mbox{\boldmath $y$}$.
Since in general the functional (\ref{gendef}) involves quark states 
of all angular momenta ($l$) a technical difficulty arises because 
this angular dependence has to be treated numerically. The major 
purpose of the present paper is to demonstrate that polarized 
structure functions can indeed be computed from a chiral soliton. 
With this in mind we will adopt the valence quark approximation 
where the quark configurations in (\ref{gendef}) are restricted to the 
valence quark level. Accordingly the valence quark wave--function 
(\ref{valrot}) is substituted into eq. (\ref{stpnt}). Then only quark 
orbital angular momenta up to $l=2$ are relevant. From a physical 
point of view this approximation is justified for moderate constituent 
quark masses ($m\approx400{\rm MeV}$) because in that parameter 
region the soliton properties are dominated by their valence quark 
contributions \cite{Al96}, \cite{Gok96}.  In particular this is the 
case for the axial properties of the nucleon. 

In the next step the polarized structure functions,  $g_1(x,\mu^2)$ 
and $g_T(x,\mu^2)$, are extracted according to eqs. (\ref{proj1}) 
and (\ref{projT}).  In the remainder of this section we will omit
explicit reference to the scale $\mu^2$.
We choose the frame such that the nucleon is
polarized along the
positive--$\mbox{\boldmath $z$}$ and positive--$\mbox{\boldmath $x$}$ 
directions in the longitudinal and transverse cases, respectively.
Note also that this implies
the choice ${\mbox{\boldmath $q$}}=q\hat{\mbox{\boldmath $z$}}$.
When extracting the structure functions the integrals 
over the time coordinate in eq. (\ref{stpnt}) can readily be done yielding the conservation 
of energy for forward and backward moving intermediate quarks. Carrying 
out the integrals over $k_0$ and $k=|\mbox{\boldmath $k$}|$ gives for 
the structure functions
\be
\hspace{-1cm}
g_1(x)&=&-N_C\frac{M_N}{\pi} 
\langle N,\frac{1}{2}\hat{\mbox{\boldmath $z$}}|\int d\Omega_{\mbox{\boldmath
$k$}} k^2
\Bigg\{\tilde\psi_{\rm v}^{\dag}(\mbox{\boldmath $p$})
\left(1-\mbox{\boldmath $\alpha$}\cdot
{\hat{\mbox{\boldmath $k$}}}\right)\gamma^5\Gamma
\tilde\psi_{\rm v}(\mbox{\boldmath $p$})
\Big|_{k=q_0+\epsilon_{\rm v}}
\nonumber \\* && \hspace{3cm}
+\tilde\psi_{\rm v}^{\dag}(-\mbox{\boldmath $p$})
\left(1-\mbox{\boldmath $\alpha$}\cdot
{\hat{\mbox{\boldmath $k$}}}\right)\gamma^5\Gamma
\tilde\psi_{\rm v}(-\mbox{\boldmath $p$})
\Big|_{k=q_0-\epsilon_{\rm v}}
\Bigg\} |N,\frac{1}{2}\hat{\mbox{\boldmath $z$}}\rangle\ ,
\label{valg1}\\
\hspace{-1cm}
g_{T}(x)&=&g_1(x)+g_2(x) 
\nonumber \\*
&=&-N_C\frac{M_N}{\pi} \langle N,\frac{1}{2}\hat{\mbox{\boldmath $x$}} |
\int d\Omega_{\mbox{\boldmath $k$}} k^2
\Bigg\{\tilde\psi_{\rm v}^{\dag}(\mbox{\boldmath $p$})
\left(\mbox{\boldmath $\alpha$}\cdot
{\hat{\mbox{\boldmath $k$}}}\right)\gamma^5\Gamma 
\tilde\psi_{\rm v}(\mbox{\boldmath $p$})
\Big|_{k=q_0+\epsilon_{\rm v}}
\nonumber \\* && \hspace{3cm}
+\tilde\psi_{\rm v}^{\dag}(-\mbox{\boldmath $p$})
\left(\mbox{\boldmath $\alpha$}\cdot
{\hat{\mbox{\boldmath $k$}}}\right)\gamma^5\Gamma 
\tilde\psi_{\rm v}(-\mbox{\boldmath $p$})
\Big|_{k=q_0-\epsilon_{\rm v}}
\Bigg\} |N,\frac{1}{2}\hat{\mbox{\boldmath $x$}} \rangle\ ,
\label{valgt}
\ee
where $\mbox{\boldmath $p$}=\mbox{\boldmath $k$}+\mbox{\boldmath $q$}$  
and $\Gamma =\frac{5}{18}{\ID} +\frac{1}{6}D_{3i}\tau_{i}$
with $D_{ij}=\frac{1}{2}\
{\rm tr}\left(\tau_{i}A(t)\tau_{j}A^{\dagger}\right)$ being the 
adjoint representation of the collective 
rotation {\it cf.} eq. (\ref{collrot}).  
The second entry 
in the states labels the spin orientation.  
$N_C$ appears as a multiplicative factor 
because the functional trace (\ref{gendef1}) includes the color 
trace as well.  Furthermore the Fourier transform of the 
valence quark wave--function
\be
\tilde\psi_{\rm v}(\mbox{\boldmath $p$})=\int \frac{d^3x}{4\pi}\
\psi_{\rm v}(\mbox{\boldmath $x$})\
{\rm exp}\left(i\mbox{\boldmath $p$}\cdot
\mbox{\boldmath $x$}\right)
\label{ftval}
\ee
has been introduced. Also, note  that the wave--function $\psi_{\rm v}$
contains an implicit dependence on the collective coordinates through
the angular velocity $\mbox{\boldmath $\Omega$}$, {\it cf.} 
eq. (\ref{valrot}).

The dependence of the wave--function 
$\tilde\psi(\pm\mbox{\boldmath $p$})$ on the integration variable 
${\hat{\mbox{\boldmath $k$}}}$ is only implicit. 
In the Bjorken 
limit the integration variables may then be changed to \cite{Ja75}
\be
k^2 \ d\Omega_{\mbox{\boldmath $k$}} =
p dp\ d\Phi\ , \qquad p=|\mbox{\boldmath $p$}|\ ,
\label{intdp}
\ee
where $\Phi$ denotes the azimuth--angle between
$\mbox{\boldmath $q$}$ and $\mbox{\boldmath $p$}$. 
The lower bound for the $p$--integral is adopted when 
$\mbox{\boldmath $k$}$ and $\mbox{\boldmath $q$}$ are anti--parallel; 
$p^{\rm min}_\pm=|M_N x\mp \epsilon_{\rm v}|$ 
for $k=-\left(q_0\pm\epsilon_{\rm v}\right)$, 
respectively.    Since the wave--function
$\tilde\psi(\pm\mbox{\boldmath $p$})$ acquires its dominant 
support for $p\le M_N$ the integrand is different from 
zero only when $\mbox{\boldmath $q$}$ and $\mbox{\boldmath $k$}$
are anti--parallel. We may therefore take 
${\hat{\mbox{\boldmath $k$}}}=-{\hat{\mbox{\boldmath $z$}}}$.
This is nothing but the light--cone description for computing 
structure functions \cite{Ja91}.  Although expected, this result is
non--trivial and will only come out in models which have a current
operator which, as in QCD, is formally identical to the one of 
non--interacting quarks. The valence quark state possesses positive 
parity yielding
$\tilde\psi(-\mbox{\boldmath $p$})=\gamma_0
\tilde\psi(\mbox{\boldmath $p$})$. 
With this we arrive at the expression for the isoscalar 
and isovector parts of the 
polarized structure function in the valence quark approximation,
\be
\hspace{-.5cm}
g^{I=0}_{1,\pm}(x)&=&-N_C\frac{5\ M_N}{18\pi}
\langle N,\frac{1}{2}\hat{\mbox{\boldmath $z$}}|
\int^\infty_{M_N|x_\mp|}p dp \int_0^{2\pi}d\Phi\
\nonumber \\* && \hspace{4cm}\times
\tilde\psi_{\rm v}^{\dag}(\mbox{\boldmath $p$}_\mp)
\left(1\pm\alpha_3\right)\gamma^5\tilde\psi_{\rm v}(\mbox{\boldmath $p$}_\mp)
|N,\frac{1}{2}\hat{\mbox{\boldmath $z$}}\rangle 
\label{g10} \\
\hspace{-.5cm}
g^{I=1}_{1,\pm}(x)&=&-N_C\frac{M_N}{6\pi}
\langle N,\frac{1}{2}\hat{\mbox{\boldmath $z$}}| D_{3i}
\int^\infty_{M_N|x_\mp|}p dp \int_0^{2\pi}d\Phi\
\nonumber \\* && \hspace{4cm}\times
\tilde\psi_{\rm v}^{\dag}(\mbox{\boldmath $p$}_\mp)\tau_i
\left(1\pm\alpha_3\right)\gamma^5\tilde\psi_{\rm v}(\mbox{\boldmath $p$}_\mp)
|N,\frac{1}{2}\hat{\mbox{\boldmath $z$}}\rangle\ ,
\label{g11}\\ 
\hspace{-.5cm}
g^{I=0}_{T,\pm}(x)&=&-N_C\frac{5\ M_N}{18\pi}
\langle N,\frac{1}{2}\hat{\mbox{\boldmath $x$}}|
\int^\infty_{M_N|x_\mp|}p dp \int_0^{2\pi}d\Phi\
\nonumber \\* && \hspace{4cm}\times
\tilde\psi_{\rm v}^{\dag}(\mbox{\boldmath $p$}_\mp)
\alpha_3\gamma^5\tilde\psi_{\rm v}(\mbox{\boldmath $p$}_\mp)
|N,\frac{1}{2}\hat{\mbox{\boldmath $x$}}\rangle 
\label{gt0}\ , \\
\hspace{-.5cm}
g^{I=1}_{T,\pm}(x)&=&-N_C\frac{M_N}{6\pi}
\langle N,\frac{1}{2}\hat{\mbox{\boldmath $x$}}| D_{3i}
\int^\infty_{M_N|x_\mp|}p dp \int_0^{2\pi}d\Phi\
\nonumber \\* && \hspace{4cm}\times
\tilde\psi_{\rm v}^{\dag}(\mbox{\boldmath $p$}_\mp)\tau_i
\alpha_3\gamma^5\tilde\psi_{\rm v}(\mbox{\boldmath $p$}_\mp)
|N,\frac{1}{2}\hat{\mbox{\boldmath $x$}}\rangle\ ,
\label{gt1} 
\ee
where $x_{\pm}=x\pm\epsilon_{\rm v}/{M_N}$ and 
${\rm cos}(\Theta^\pm_p)={M_N}x_\pm/{p}$.
The complete structure functions are given by
\be
g_{1}(x)&=&g^{I=0}_{1,+}(x)+g^{I=1}_{1,+}(x)
-\left(g^{I=0}_{1,-}(x)-g^{I=1}_{1,-}(x)\right)
\label{gone}  \\*  \hspace{-1cm} 
g_{T}(x)&=&g^{I=0}_{T,+}(x)+g^{I=1}_{T,+}(x)
-\left(g^{I=0}_{T,-}(x)-g^{I=1}_{T,-}(x)\right)\ .
\label{gtran}
\ee 
Note also, that we have made explicit the isoscalar
$\left(I=0\right)$
and isovector $\left(I=1\right)$ parts.
The wave--function  implicitly depends
on $x$ because 
$\tilde\psi_{\rm v}(\mbox{\boldmath $p$}_\pm)=
\tilde\psi_{\rm v}(p,\Theta^\pm_p,\Phi)$
where the polar--angle, $\Theta^\pm_p$, between $\mbox{\boldmath $p$}_\pm$ 
and $\mbox{\boldmath $q$}$ is fixed for a given value of the Bjorken 
scaling variable $x$. 

Turning to the evaluation of the nucleon matrix elements defined 
above we first note that the Fourier transform of the wave--function 
is easily obtained because the angular parts are tensor spherical 
harmonics in both coordinate and momentum spaces. Hence, only the 
radial part requires numerical treatment. 
Performing straightforwardly
the azimuthal integrations in eqs. (\ref{g10}) and (\ref{g11})
reveals that the surviving isoscalar part of the longitudinal structure
function, $g_{1}^{I=0}$, is linear in the angular velocity,
$\mbox{\boldmath $\Omega$}$. It is this part which is associated  with the
proton--spin puzzle.  Using the standard quantization condition,
$\mbox{\boldmath $\Omega$} =\mbox{\boldmath $J$}/\ \alpha^2$, 
where $\alpha^2$ is the moment of inertia of the soliton
and further noting that 
the ${\hat{\mbox{\boldmath $z$}}}$--direction is distinct, 
the required nucleon matrix elements are 
$\langle N,\frac{1}{2}\hat{\mbox{\boldmath $z$}}|
J_{z}|N,\frac{1}{2}\hat{\mbox{\boldmath $z$}}\rangle=\frac{1}{2}$.
Thus, $g_1^{I=0}$ is identical for all nucleon states. 
Choosing a symmetric  ordering \cite{Al93}, \cite{Sch95} for 
the non--commuting operators, 
$D_{ia}J_j\rightarrow \frac{1}{2}\left\{D_{ia},J_j \right\}$
we find that the nucleon matrix elements associated with the 
cranking portion of  the isovector piece, $\langle
N,\pm\frac{1}{2}\hat{\mbox{\boldmath $z$}}|\left\{D_{3y},J_x
\right\}|N,\pm\frac{1}{2}\hat{\mbox{\boldmath $z$}}\rangle$, vanish.
With this ordering we avoid the occurance of PCAC violating pieces in
the axial current.  
The surviving terms stem solely from the classical part of the
valence quark wave--function, 
$\Psi_{\rm v}\left({\mbox{\boldmath $x$}}\right)$  in
combination with the collective Wigner--D function, $D_{3z}$.  Again
singling out the ${\hat{\mbox{\boldmath $z$}}}$--direction, 
the nucleon matrix elements become \cite{Ad83}
\be
\langle
N,\frac{1}{2}\hat{\mbox{\boldmath $z$}}|D_{3z}|N,\frac{1}{2}\hat{\mbox{\boldmath $z$}}
\rangle = -\frac{2}{3} i_3\ ,
\label{matz}
\ee
where $i_3=\pm\frac{1}{2}$ is the nucleon isospin.
For the transverse structure function, the surviving piece of the
isoscalar contribution is again linear in the angular velocities.
The transversally polarized nucleon gives rise to
the matrix elements, 
$\langle N,\frac{1}{2}\hat{\mbox{\boldmath $x$}}
|J_{x}|N,\frac{1}{2}\hat{\mbox{\boldmath $x$}}\rangle=\frac{1}{2}$.  
Again choosing symmetric ordering for terms
arising from the cranking contribution, the nucleon matrix elements
$\langle
N,\frac{1}{2}\hat{\mbox{\boldmath $x$}}|\left\{D_{3y},J_y
\right\}|N,\frac{1}{2}\hat{\mbox{\boldmath $x$}}\rangle$
and $\langle
N,\frac{1}{2}\hat{\mbox{\boldmath $x$}}|\left\{D_{33},J_y
\right\}|N,\frac{1}{2}\hat{\mbox{\boldmath $x$}}\rangle$ vanish.
As in the longitudinal case, there is a surviving isovector contribution 
stemming solely from the classical part of the
valence quark wave--function, $\Psi_{\rm v}({\mbox{\boldmath $x$}})$ 
in combination with the collective Wigner--D function, $D_{3x}$.  
Now singling out the $\hat{\mbox{\boldmath $x$}}$--direction 
the relevant nucleon matrix elements become \cite{Ad83},
\be
\langle
N,\frac{1}{2}\hat{\mbox{\boldmath $x$}}|D_{3x}
|N,\frac{1}{2}\hat{\mbox{\boldmath $x$}}\rangle
= -\frac{2}{3} i_3\  .
\label{matx}
\ee
Explicit expressions  in terms of the valence quark
wave functions (\ref{gone} and \ref{gtran}) for 
$g^{I=0}_{1,\pm}(x)$, $g^{I=1}_{1,\pm}(x)$, $g^{I=0}_{2,\pm}(x)$ 
and $g^{I=1}_{,\pm}(x)$ are listed in
the appendix A.

Using the expressions given in the appendix A
it is straightforward to verify the Bjorken sum rule \cite{Bj66} 
\be
\Gamma_1^{p}-\Gamma_1^{n}&=&\int_{0}^{1} dx\ \left(g_{1}^{p}(x)-
g_{1}^{n}(x)\right)=g_{A}/6\ ,
\label{bjs}
\ee
the Burkhardt--Cottingham sum rule \cite{bur70} 
\be
\Gamma_2^{p}&=&\int_{0}^{1} dx\ g_{2}^{p}(x)=0\ ,
\label{bcs}   
\ee
as well as the axial singlet charge
\be
\Gamma_1^{p}+\Gamma_1^{n}&=&\int_{0}^{1} dx\ \left(g_{1}^{p}(x)+
g_{1}^{n}(x)\right)=g_A^{0}\ ,
\label{gas}
\ee
in this model calculation when the moment of inertia 
$\alpha^2$, as well as the axial charges $g_A^0$ and $g_A$, are 
confined to their dominating valence quark pieces.
We have used
\be
g_A&=&-\frac{N_C}{3}\int d^3 r 
{\bar\psi}_{\rm v}^{\dagger}(\mbox{\boldmath $r$})\gamma_3
\gamma_5\tau_3 \psi_{\rm v}(\mbox{\boldmath $r$})
\label{gaval} \\
g_A^0&=&\frac{N_C}{\alpha_{\rm v}^2}
\int d^3 r{\bar\psi}_{\rm v}^{\dagger}(\mbox{\boldmath $r$})\gamma_3
\gamma_5\psi_{\rm v}(\mbox{\boldmath $r$}) \ .
\label{ga0val}
\ee
to verify the Bjorken Sum rule as well as the axial singlet charge.
This serves as an analytic check on our treatment.
Here $\alpha_{\rm v}^2$ refers to the valence quark contribution
to the moment of inertia, {\it i.e.}
$\alpha_{\rm v}^2=(1/2)\sum_{\mu\ne{\rm v}}
|\langle\mu|\tau_3|{\rm v}\rangle|^2/(\epsilon_\mu-\epsilon_{\rm v})$.
The restriction to the valence quark piece is required by consistency 
with the Adler sum rule in the calculation of the unpolarized 
structure functions in this approximation \cite{wei96}.

\bigskip
\section{Numerical Results}
\bigskip

In this section we display the results of the spin--polarized
structure functions calculated from eqs. (\ref{g1zro}--\ref{gton}) 
for constituent quark masses of $m=400{\rm MeV}$ and $450{\rm MeV}$.  
In addition to checking the above mentioned  sum rules  
see eqs. (\ref{bjs})--(\ref{gas}), 
we have numerically calculated the 
first moment of $g_{1}^{p}(x,\mu^{2})$\footnote{Which in
this case amounts to the  Ellis--Jaffe sum rule \cite{Ja74} 
since we have omitted the strange degrees of freedom. A careful
treatment of symmetry breaking effects indicates that the role of the
strange quarks is 
less important than originally assumed \cite{jon90,Li95}.}
\be
\Gamma_1^{p}&=&\int_{0}^{1} dx\ g_{1}^{p}(x)\ ,
\label{ejs} 
\ee
and the
Efremov--Leader--Teryaev (ELT) sum rule \cite{Ef84}
\be
\Gamma_{\rm ETL}&=&\int_{0}^{1} dx\ x\left(g_{1}^{p}(x) 
+2g_{2}^{n}(x)\right)\ .
\label{elts}
\ee
We summarize the results for the sum rules in table 1.
When comparing these results with the experimental data one observes
two short--comings, which are already known from studies of the 
static properties in this model. First, the axial charge 
$g_A\approx 0.73$ comes 
out too low as the experimental value is $g_A=1.25$. It has 
recently been speculated that a different ordering of the collective 
operators $D_{ai}J_j$ ({\it cf.} section 4) may fill the gap
\cite{Wa93,Gok96}. However, since such an ordering unfortunately gives 
rise to PCAC violating contributions to the axial current \cite{Al93}
and furthermore inconsistencies with $G$--parity may occur in 
the valence quark approximation \cite{Sch95} we will not pursue
this issue any further at this time. Second, the predicted axial singlet 
charge $g_A^0\approx 0.6$ is approximately twice as large 
as the number extracted from experiment\footnote{Note 
that this analysis assumes $SU(3)$ flavor symmetry, which, of course, 
is not manifest in our two flavor model.} $0.27\pm0.04$\cite{ell96}. 
This can be 
traced back to the valence quark approximation as there are direct 
and indirect contributions to $g_A^0$ from both the polarized 
vacuum and the valence quark level. Before canonical quantization 
of the collective coordinates one finds a sum of valence 
and vacuum pieces
\be
g_A^0=2\left(g_{\rm v}^0+g_{\rm vac}^0\right)\Omega_3
=\frac{g_{\rm v}^0+g_{\rm vac}^0}
{\alpha^2_{\rm v}+\alpha^2_{\rm vac}} \ .
\label{ga0val1}
\ee
Numerically the vacuum piece is negligible, {\it i.e.}
$g_{\rm vac}^0/g_{\rm v}^0\approx 2\%$. Canonical quantization 
subsequently involves the moment of inertia 
$\alpha^2=\alpha^2_{\rm v}+\alpha^2_{\rm vac}$, which also has 
valence and vacuum pieces. In this case, however, the vacuum 
part is not so small: $\alpha^2_{\rm vac}/\alpha^2\approx25\%$.
Hence the full treatment of the polarized vacuum will drastically 
improve the agreement with the empirical value for $g_A^0$.
On the other hand our model calculation nicely reproduces the 
Ellis--Jaffe sum rule since the empirical value is $0.136$.
Note that this comparison is legitimate since neither the 
derivation of this sum rule nor our model imply strange quarks.
While the vanishing Burkhardt--Cottingham sum rule can be 
shown analytically in this model, the small value for the 
Efremov--Leader--Teryaev sum rule is a numerical prediction.
Recently, it has  been demonstrated \cite{So94} that that the ELT 
sum rule (\ref{elts}), which is derived within the parton model,
neither vanishes in the Center of Mass bag model\cite{So94} 
nor is supported by the
SLAC E143 data \cite{slac96}.  This is also the case for our
NJL--model calculation as can be seen from table I.  

In figure 1 we display the spin structure functions 
$g_{1}^{p}(x,\mu^{2})$ and $g_{2}^{p}(x,\mu^{2})$ along with the
twist--2 piece, $g_{2}^{WW(p)}\left(x,\mu^{2}\right)$ and twist--3
piece, ${\overline{g}}_{2}^{p}\left(x,\mu^{2}\right)$. The actual 
value for $\mu^2$ will be given in the proceeding section in the 
context of the evolution procedure.  We observe that the structure 
functions $g_{2}^{p}(x,\mu^{2})$ and $g_{2}^{WW(p)}(x,\mu^{2})$ are 
well localized in the interval $0\le x\le1$, while for $g_1^{p}$ about 
$0.3\%$ of the first moment, 
$\Gamma_1^{p}=\int_{0}^{1} dx\ g_{1}^{p}(x,\mu^2)$ 
comes from the region, $x > 1$.  
The polarized structure function $g_1^{p}(x,\mu^2)$ exhibits a 
pronounced maximum at $x\approx0.3$ which is smeared out when the 
constituent quark mass increases.  This can be understood as follows:
In our chiral soliton model the constituent mass serves as a coupling
constant of the quarks to the chiral field (see eqs. (\ref{bosact})
and (\ref{hamil})).  
The valence quark
becomes more strongly bound as the constituent quark mass increases. 
In this case the lower components of the valence quark 
wave--function increase and relativistic effects 
become more important resulting in a broadening of the maximum.
With regard to the Burkhardt--Cottingham sum rule the polarized
structure function $g_2^{p}(x,\mu^2)$ possesses a node. Apparently 
this node appears at approximately the same value of the Bjorken 
variable $x$ as the maximum of $g_1^{p}(x,\mu^2)$. Note also that 
the distinct twist contributions to $g_2^{p}(x,\mu^2)$ 
by construction diverge as ${\rm ln}\left(x\right)$ as 
$x\to0$ while their sum stays finite(see section 6 for details).

As the results displayed in figure 1 are the central issue of 
our calculation it is of great interest to compare them with the 
available data. As for all effective low--energy models of the
nucleon, the predicted results are at a lower scale $Q^2$ than 
the experimental data. In order to carry out a sensible comparison 
either the model results have to be evolved upward or the QCD 
renormalization group equations have to be used to extract structure 
functions at a low--renormalization point. For the combination 
$xg_1(x)$ a parametrization of the empirical structure function 
is available at a low scale \cite{Gl95}\footnote{These authors also 
provide a low scale parametrization of quark distribution functions. 
However, these refer to the distributions of perturbatively interacting 
partons. Distributions for the NJL--model constituent quarks could 
in principle be extracted from eqs. (\ref{g10})--(\ref{gt1}). It is 
important to stress that these distributions may not be compared 
to those of ref \cite{Gl95} because the associated quarks fields 
are different in nature.}. In that study the experimental high $Q^2$ 
data are evolved to the low--renormalization point $\mu^2$, which is 
defined as the lowest $Q^2$ satisfying the positivity constraint 
between the polarized and unpolarized structure functions.  In a 
next--to--leading order calculation those authors found 
$\mu^2=0.34{\rm GeV}^2$ \cite{Gl95}. In figure 2 we compare our 
results for two different constituent quark masses with that 
parametrization. We observe that our predictions reproduce gross 
features like the position of the maximum. This agreement is the 
more pronounced the lower the constituent quark is, {\it i.e.} the 
agreement improves as the applicability of the valence quark 
approximation becomes more justified. Unfortunately, such a 
parametrization is currently not available for the transverse 
structure function $g_T(x)$ (or $g_2(x)$). In order to nevertheless 
be able to compare our corresponding results with the (few) available 
data we will apply leading order evolution techniques to the structure 
functions calculated in the valence quark approximation to the 
NJL--soliton model. This will be subject of the following section.

\bigskip
\section{Projection and Evolution}
\bigskip

One notices that our baryon states are not momentum eigenstates 
causing the structure functions (see figures 1 and 2) 
not to vanish exactly for $x>1$ although the contributions
for $x>1$ are very small.  This short--coming is due to the 
localized field configuration and thus the nucleon not being a 
representation of the Poincar\'{e} group which is common to the 
low--energy effective models. The most feasible procedure to cure 
this problem is to apply Jaffe's prescription \cite{Ja80},  
\be
f(x)\longrightarrow \tilde f(x)=
\frac{1}{1-x}f\left(-{\rm log}(1-x)\right)
\label{proj}
\ee
to project any structure function $f(x)$ onto the interval
$[0,1]$.  In view of the kinematic regime of DIS this
prescription, which was
derived in a Lorentz invariant fashion within the 1+1 dimensional 
bag model, is a reasonable approximation.  It is important to
note in the NJL model the unprojected nucleon wave--function
(including the cranking piece\footnote{Which in fact yields the 
leading order to the Adler sum rule, 
$F_1^{\nu p}\ - F_1^{{\bar \nu}p}$ \cite{wei96} rather than being a 
correction.}, see \ref{valrot}) is anything but a product of
Dirac--spinors.  In this context, techniques such as  
Peierls--Yoccoz\cite{Pei57} (which does not completely enforce 
proper support \cite{Sig90}, $0\le x\le1$ nor restore Lorentz 
invariance, see \cite{Ard93}) appear to be infeasible.  Thus, 
given the manner in which the nucleon
arises in chiral--soliton models Jaffe's projection 
technique is quite well suited.  
It is also important to note
that, by construction, sum rules are not effected by this 
projection, {\it i.e.}
$\int_0^\infty dxf(x)=
\int_0^1 dx \tilde f(x)$.  Accordingly the sum--rules of the
previous section remain intact.

With regard to evolution of the spin--polarized structure functions
applying the OPE analysis of Section 2, Jaffe and Ji brought to light
that to leading order in $1/Q^{2}$,
$g_1(x,Q^2)$ receives only a leading order twist--2 
contribution, while $g_2(x,Q^2)$ possesses contributions 
from both twist--2 and twist--3 operators; 
the twist--3 portion coming from
spin--dependent gluonic--quark correlations \cite{Ja90},\cite{Ji90}
(see also, \cite{ko79} and \cite{sh82}).
In the {\em impulse approximation} 
\cite{Ja90}, \cite{Ji90}
these leading contributions are given by
\be
\hspace{-2cm}
\lim_{Q^2\to\infty}
\int_{0}^{1} dx\ x^{n} g_{1}(x,Q^2)&=&\frac{1}{2}\sum _{i}\
{\cal O}_{2,i}^{n}\ \ ,\ \ n=0,2,4,\ldots\ ,
\label{ltc1} \\
\lim_{Q^2\to\infty}
\int_{0}^{1} dx\ x^{n}\ g_{2}(x,Q^2)&=&-\frac{n}{2\ (n+1)} 
\sum_{i} \left\{ {\cal O}_{2,i}^{n} 
-{\cal O}_{3,i}^{n} \right\},\  n=2,4,\ldots\ .
\label{ltc2}
\ee
Note that there is no sum rule for the first
moment, $\Gamma_{2}(Q^2)=\int_{0}^{1}\ dx g_{2}(x,Q^2)$, \cite{Ja90}.  
Sometime ago Wandzura and Wilczek \cite{wan77}
proposed that $g_2(x,Q^2)$ was given in terms of $g_1(x,Q^2)$,
\be
g_{2}^{WW}(x,Q^2)=-\ g_{1}(x,Q^2)+\ \int_{x}^{1}\frac{dy}{y}\ g_{1}(y,Q^2)
\label{ww}
\ee
which follows immediately from eqs. (\ref{ltc1}) and (\ref{ltc2})
by neglecting the twist--3 portion in the sum in
(\ref{ltc2}).  One may reformulate 
this argument to extract the twist--3 piece
\be
{\overline{g}}_{2}(x,Q^2)\ =\ g_{2}(x,Q^2)\ -\ g_{2}^{WW}(x,Q^2)\ ,
\ee
since,
\be
\int_{0}^{1} dx\ x^{n}\ {\overline{g}}_{2}(x,Q^2)=\frac{n}{2\ (n+1)} 
\sum_{i} {\cal O}_{3,i}^{n}\ \ ,  \ n=2,4,\ \ldots \ .
\ee

In the NJL model as in the bag--model there are no explicit gluon
degrees of freedom, however, in both models twist--3
contributions to $g_2(x,\mu^2)$ exist.  In contrast to the bag
model where the bag boundary simulates the quark--gluon and
gluon--gluon correlations  \cite{So94} in the NJL model the 
gluon degrees of freedom, having been ``integrated" out, 
leave correlations characterized by the four--point quark 
coupling $G_{\rm NJL}$.  This is the source of the twist--3
contribution to $g_2(x,\mu^2)$, which is shown in figure 1.

For $g_{1}\left(x,Q^2\right)$ and the twist--2 piece
$g_2^{WW}\left(x,Q^2\right)$
we apply the leading 
order (in $\alpha_{QCD}(Q^2)$) Altarelli--Parisi 
equations \cite{Al73} to evolve 
the structure functions from the model 
scale, $\mu^2$,  to that 
of the experiment $Q^2$, by iterating
\be
g(x,t+\delta{t})=g(x,t)\ +\ \delta t\frac{dg(x,t)}{dt}\ ,
\ee
where $t={\rm log}\left(Q^2/\Lambda_{QCD}^2\right)$.
The explicit expression for the evolution differential 
equation is given by the convolution integral,
\be
\frac{d g(x,t)}{dt}&=&\frac{\alpha(t)}{2\pi}
g(x,t)\otimes P_{qq}(x)
\nonumber  \\*  \hspace{1cm} 
&=&\frac{\alpha(t)}{2\pi}
C_{R}(F)\int^1_{x}\ \frac{dy}{y}P_{qq}\left(y\right)
g\left(\frac{x}{y},t\right)
\label{convl}
\ee
where the quantity 
$P_{qq}\left(z\right)=\left(\frac{1+z^2}{1-z^2}\right)_{+}$ 
represents the quark probability to emit a gluon such that the 
momentum of the quark is reduced by the fraction $z$.
$C_{R}(f)=\frac{{n_{f}^{2}}-1}{2{n_{f}}}$ for $n_f$--flavors,
$\alpha_{QCD}=\frac{4\pi}{\beta\log\left(Q^2/ \Lambda^2\right)}$
and $\beta=(11-\frac{2}{3}n_f)$.
Employing the ``+" prescription\cite{Al94} yields
\be
\frac{d\ g(x,t)}{dt}&=&\frac{2C_{R}(f)}{9\ t}
\left\{\ \left(x + \frac{x^{2}}{2}+2\log(1-x)\right)g(x,t)
\right.
\nonumber  \\*&& \hspace{1cm} 
\left. 
+\ \int^{1}_{x}\ dy
\left(\frac{1+y^2}{1-y}\right)
\left[\frac{1}{y}\ g\left(\frac{x}{y},t\right)-g(x,t)\right]\
\right\}\ .
\label{evol}
\ee
As discussed in section 2 the initial value for integrating the
differential equation is given by the scale $\mu^2$ at which the model is
defined.  It should be emphasized that this scale essentially is a 
new parameter of the model.  For a given constituent quark mass we fit
$\mu^2$ to maximize the agreement of the predictions with the
experimental data on previously \cite{wei96} calculated unpolarized
structure functions for (anti)neutrino--proton scattering: 
$F_2^{\nu p}-F_2^{\overline{\nu} p}$. For the constituent quark mass 
$m=400{\rm MeV}$ we have obtained $\mu^2\approx0.4{\rm GeV}^2$.
One certainly wonders whether for such a low scale the restriction to 
first order in $\alpha_{QCD}$ is reliable. There are two answers. First,
the studies in this section aim at showing that the required evolution 
indeed improves the agreement with the experimental data and, second,
in the bag model it has recently been shown \cite{St95} that a 
second order evolution just increases $\mu^2$ without significantly 
changing the evolved data.  In figure 3 we compare the 
unevolved, projected, structure function
$g_1^{p}\left(x,\mu^{2}\right)$ with the one 
evolved from $\mu^{2}=0.4{\rm GeV}^2$ to $Q^2=3.0{\rm GeV}^2$.
Also the data from the E143-collaboration from
SLAC\cite{slac95a} are given. Furthermore in 
figure 3 we compare the projected, unevolved structure 
function $g_2^{WW(p)}\left(x,\mu^{2}\right)$ as well as the one evolved 
to $Q^2=5.0{\rm GeV}^2$ with the data from the recent E143-collaboration 
at SLAC\cite{slac96}.
As expected we observe that the evolution pronounces the structure
function at low $x$; thereby improving the agreement with the 
experimental data. This change towards small $x$ is a general feature 
of the projection and evolution process and presumably not very 
sensitive to the prescription applied here. In particular, choosing 
an alternative projection technique may easily be compensated by 
an appropriate variation of the scale $\mu^2$.

While the evolution of the structure function
$g_{1}\left(x,Q^2\right)$ and the twist--2 piece
$g_2^{WW}\left(x,Q^2\right)$ from $\mu^2$ to $Q^2$ can be performed 
straightforwardly using the ordinary Altarelli--Parisi equations
this is not the case with the twist--3 piece
${\overline{g}}_{2}(x,Q^2)$.   
As the twist--3 quark and quark--gluon operators mix
the number of independent operators contributing 
to the twist--3 piece increases 
with $n$, where $n$ refers to the $n^{\underline{\rm th}}$ moment\cite{sh82}.  
We apply an approximation (see appendix B) suggested in\cite{Ali91}
where it is demonstrated
that in  $N_c\to \infty$ limit the quark
operators of twist--3 decouple from the evolution equation
for the quark--gluon operators of the same twist resulting
in a unique evolution scheme.
This scheme is particularly suited for the NJL--chiral soliton model,
as the soliton picture for baryons is based on $N_c\rightarrow \infty$
arguments\footnote{This scheme has also employed by Song\cite{So94} 
in the Center of Mass bag model.}.

In figure 4 we compare the projected unevolved structure function
${\overline{g}}_{2}^{p}(x,\mu^2)$ evolved to $Q^2=5.0{\rm GeV}^2$
using the scheme suggested in \cite{Ali91}.  In addition 
we reconstruct $g_2^{p}\left(x,Q^2\right)$ at $Q^2=3.0{\rm GeV}^2$ from
$g_2^{WW(p)}\left(x,Q^2\right)$ and ${\overline{g}}_{2}(x,Q^2)$
and compare it with the recent SLAC data\cite{slac96}
for $g_2^{p}\left(x,Q^2\right)$.  As is evident our model
calculation of $g_2^{p}\left(x,Q^2\right)$,
built up from its twist--2 and twist--3 pieces, 
agrees reasonably well
with the data although the experimental errors are quite large.

\bigskip
\section{Summary and Outlook}
\bigskip
In this paper we have presented the calculation of the polarized 
nucleon structure functions $g_1\left(x,Q^2\right)$ and
$g_2\left(x,Q^2\right)$  within a model which is 
based on chiral symmetry and its spontaneous breaking. Specifically
we have employed the NJL chiral soliton model which reasonably 
describes the static properties of the nucleon \cite{Al96},
\cite{Gok96}.  In this model the 
current operator is formally identical to the one in an
non--interacting relativistic quark model. While the quark 
fields become functionals of the chiral soliton upon bosonization,
this feature enables one calculate the hadronic tensor.
From this hadronic tensor we have then extracted the polarized 
structure functions within the valence quark approximation. As the 
explicit occupation of the valence quark level yields the 
major contribution (about 90\%) to the associated static quantities 
like the axial charge this presumably is a justified approximation. 
When cranking corrections are included this share may be reduced 
depending on whether or not the full moment of inertia is substituted.

It needs to be stressed that in contrast to {\it e.g.} bag models 
the nucleon wave--function arises as a collective excitation of 
a non--perturbative meson field configuration. In particular, the 
incorporation of chiral symmetry leads to the distinct feature that 
the pion field cannot be treated perturbatively. Because of the
hedgehog structure of this field one starts with grand spin symmetric
quark wave--functions rather than direct products of spatial-- and 
isospinors as in the bag model. On top of these grand spin 
wave--functions one has to include cranking corrections to generate 
states with the correct nucleon quantum numbers. Not only are these 
corrections sizable but even more importantly one would not be able 
to make any prediction on the flavor singlet combination of the 
polarized structure functions without them. The structure functions 
obtained in this manner are, of course, characterized by the scale 
of the low--energy effective model. We have confirmed this issue by 
obtaining a reasonable agreement of the model predictions for the 
structure function $g_1$ of the proton with the low--renormalization 
point parametrization of ref \cite{Gl95}. In general this scale of 
the effective model essentially represents an intrinsic parameter of 
a model. For the NJL--soliton model we have previously determined
this parameter from the behavior of the unpolarized structure 
functions under the Altarelli--Parisi evolution \cite{wei96}. Applying 
the same procedure to the polarized structure functions calculated in 
the NJL model yields good agreement with the data extracted from 
experiment, although the error bars on $g_1\left(x,Q^2\right)$ are 
still sizable. In particular, the good agreement at low $x$ indicates 
that to some extend gluonic effects are already incorporated in the 
model. This can be understood by noting that the quark fields, which 
enter our calculation, are constituent quarks. They differ from the 
current quarks by a mesonic cloud which contains gluonic components. 
Furthermore, the existence of gluonic effects in the model would not 
be astonishing because we had already observed from the non--vanishing 
twist--3 part of $g_2\left(x,Q^2\right)$, which in the OPE is 
associated with the quark--gluon interaction, that the model contains 
the main features allocated to the gluons.

There is a wide avenue for further studies in this model. Of course,
one would like to incorporate the effects of the polarized vacuum, 
although one expects from the results on the static axial properties 
that their direct contributions are negligible. It may be more 
illuminating to include the strange quarks within the valence quark 
approximation. This extension of the model seems to be demanded by 
the analysis of the proton spin puzzle. Technically two changes will 
occur. First, the collective matrix elements will be more complicated than 
in eqs. (\ref{matz}) and (\ref{matx}) because the nucleon wave--functions
will be distorted $SU(3)$ $D$--functions in the presence of flavor 
symmetry breaking \cite{Ya88,wei96a}. Furthermore the valence quark 
wave--function (\ref{valrot}) will contain an additional correction 
due to different non--strange and strange constituent quark masses 
\cite{We92}. When these corrections are included direct information 
will be obtained on the contributions of the strange quarks to polarized 
nucleon structure functions. In particular the previously developed 
generalization to three flavors \cite{We92} allows one to 
consistently include the effects of flavor symmetry breaking.

\bigskip
\acknowledgements
This work is supported in part by the Deutsche 
Forschungsgemeinschaft (DFG) under contract Re 856/2-2.
LG is grateful for helpful comments by G. R. Goldstein.
\bigskip
\bigskip

\appendix
\section{Spin--Polarized Structure Functions}
\bigskip

In this appendix we summarize the explicit expressions for the
structure functions, eqs. (\ref{g10}--\ref{gt1}). The first step
is to construct the eigenfunctions of the single particle Dirac 
Hamiltonian (\ref{hamil}) in coordinate space. As the hedgehog {\it
ansatz} (\ref{hedgehog}) connects coordinate space with isospace,
these eigenfunctions are also eigenstates of the grand spin 
operator
\be
{\mbox{\boldmath $G$}}=
{\mbox{\boldmath $J$}}+\frac{{\mbox{\boldmath $\tau$}}}{2}
={\mbox{\boldmath $l$}}+\frac{{\mbox{\boldmath $\sigma$}}}{2}
+\frac{{\mbox{\boldmath $\tau$}}}{2}
\label{gspin}
\ee
which is the sum of the total spin ${\mbox{\boldmath $J$}}$ and the
isospin ${\mbox{\boldmath $\tau$}}/2$. The spin itself is decomposed
into orbital angular momentum ${\mbox{\boldmath $l$}}$ and intrinsic
spin ${\mbox{\boldmath $\sigma$}}/2$. Denoting by $M$ the grand 
spin projection quantum number the tensor spherical harmonics which 
are associated with the grand spin may be written as
${\cal Y}^{G,M}_{l,j}({\hat {\mbox{\boldmath $r$}}})$. Note that 
these tensor spherical harmonics are two--component spinors in 
both spin and isospin spaces. Given a fixed profile function 
$\Theta(r)$ the numerical diagonalization of the Dirac Hamiltonian 
(\ref{hamil}) yields the radial functions 
$g_\mu^{(G,+,1)}(r),f_\mu^{(G,+,1)}(r)$, etc. in 
({\it cf.} ref \cite{Ka84})
\be
\Psi_\mu^{(G,+)}(\mbox{\boldmath $r$})=
\pmatrix{ig_\mu^{(G,+;1)}(r){\cal Y}^{G,M}_{G,G+\frac{1}{2}} 
({\hat {\mbox{\boldmath $r$}}}) \cr
f_\mu^{(G,+;1)}(r) {\cal Y}^{G,M}_{G+1,G+\frac{1}{2}} 
({\hat {\mbox{\boldmath $r$}}})\cr} +
\pmatrix{ig_\mu^{(G,+;2)}(r){\cal Y}^{G,M}_{G,G-\frac{1}{2}} 
({\hat {\mbox{\boldmath $r$}}})\cr
-f_\mu^{(G,+;2)}(r) {\cal Y}^{G,M}_{G-1,G-\frac{1}{2}}
({\hat {\mbox{\boldmath $r$}}})\cr}
\label{psipos} \\ \\
\Psi_\mu^{(G,-)}(\mbox{\boldmath $r$})=
\pmatrix{ig_\mu^{(G,-;1)}(r) {\cal Y}^{G,M}_{G+1,G+\frac{1}{2}} 
({\hat {\mbox{\boldmath $r$}}})\cr
-f_\mu^{(G,-;1)}(r) {\cal Y}^{G,M}_{G,G+\frac{1}{2}} 
({\hat {\mbox{\boldmath $r$}}})\cr} +
\pmatrix{ig_\mu^{(G,-;2)}(r){\cal Y}^{G,M}_{G-1,G-\frac{1}{2}} 
({\hat {\mbox{\boldmath $r$}}})\cr
f_\mu^{(G,-;2)}(r) {\cal Y}^{G,M}_{G,G-\frac{1}{2}} 
({\hat {\mbox{\boldmath $r$}}})\cr}.
\label{psineg}
\ee
The second superscript ($\pm$) denotes the intrinsic parity, which 
also is a conserved quantum number\footnote{The total parity is given 
by the product of the intrinsic parity and $(-)^G$.}. Note that for 
the $G=0$ channel, which contains the classical contribution to the 
valence quark wave--function in eq. (\ref{valrot})
\be
\Psi_{\rm v}({\mbox{\boldmath $r$}})=\left(
\ba{c}
i g_{\rm v}(r) {\cal Y}_{0,\frac{1}{2}}^{0,0} 
\left(\hat{\mbox{\boldmath $r$}}\right) \\
f_{\rm v}(r) {\cal Y}_{1,\frac{1}{2}}^{0,0}
\left(\hat{\mbox{\boldmath $r$}}\right)
\ea
\right),
\label{valcl}
\ee
only the components with $j=+1/2$ are allowed. In addition to 
the classical piece (\ref{valcl}) the complete valence quark 
wave--function (\ref{valrot}) also contains the cranking correction, 
which dwells in the channel with $G=1$ and negative intrinsic parity.

The discretization ($\mu$) is accomplished by choosing suitable 
boundary conditions at a radial distance which is large compared to 
the soliton extension \cite{Ka84,We92}. This calculation yields the 
energy eigenvalues $\epsilon_\mu$, which enter the energy functional 
(\ref{efunct}). The soliton configuration is finally determined by 
self--consistently minimizing this energy functional. In ref. 
\cite{Al94a} the numerical procedure is described in detail.

We continue by making explicit the Fourier transform (\ref{ftval}) of 
eq. (\ref{valrot}),
\be
\tilde{\psi}_{\rm v}\left(\mbox{\boldmath $p$}\right)=\tilde{\Psi}_{\rm
v}\left(\mbox{\boldmath $p$}\right)+Q_\mu
\tilde{\Psi}_{\mu}\left(\mbox{\boldmath $p$}\right)\ .
\ee
The leading order in $N_c$ valence quark contribution is just 
Fourier transform of (\ref{valcl})
\be
\tilde\Psi_{\rm v}({\mbox{\boldmath $p$}})=i\left(
\ba{c}
\gvp\ {\cal Y}_{0,\frac{1}{2}}^{0,0} \left(\hat{\mbox{\boldmath $p$}}\right)
\\ \fvp\ {\cal Y}_{1,\frac{1}{2}}^{0,0}\left(\hat{\mbox{\boldmath $p$}}\right)
\ea
\right)
\ee
and the cranking correction involves the Fourier transform of 
spinor with $G=1$ and negative intrinsic parity
\be
\tilde\Psi_{\mu}({\mbox{\boldmath $p$}})=
-i\left(\ba{c}\guip\ {\cal Y}_{2,\frac{3}{2}}^{1,M}
\left(\hat{\mbox{\boldmath $p$}}\right) -
\guiip\ {\cal Y}_{0,\frac{1}{2}}^{1,M}
\left(\hat{\mbox{\boldmath $p$}}\right)\\ 
\fuip\ {\cal Y}_{1,\frac{3}{2}}^{1,M}
\left(\hat{\mbox{\boldmath $p$}}\right)-
\fuiip\ {\cal Y}_{1,\frac{1}{2}}^{1,M}
\left(\hat{\mbox{\boldmath $p$}}\right)
\ea
\right) \ .
\ee
Here ${\cal Y}_{l,j}^{G,M}\left(\hat{\mbox{\boldmath $p$}}\right)$ 
are the Fourier transforms of the tensor spherical harmonics 
associated with the grand spin operator (\ref{gspin}). 
The Fourier transform for the radial functions is defined by
\be
\tilde{\phi}_\mu(p)=\int_{0}^{R}dr\ r^2
j_{l}(pr)\phi_{\mu}(r)\ .
\ee
Here the index $l$ of the Bessel function
denotes the orbital angular momentum of the associated tensor 
spherical harmonic. We have suppressed the grand spin index 
on the transforms of the radial wave functions for convenience.
For purposes of notation we have introduced 
the quantity, $Q_\mu$ which
arises in analyzing the matrix elements (see eq. (\ref{valrot}))
\be
\frac{\langle \mu |\mbox{\boldmath $\tau$}\cdot
\mbox{\boldmath $\Omega$}|{\rm v}\rangle}
{\epsilon_{\rm v}-\epsilon_\mu}&=&
Q_\mu\ \left\{\frac{\delta_{M,1}}{\sqrt 2}
\left(\Omega_{1}+\Omega_{2}\right)
-\frac{\delta_{M,-1}}{\sqrt 2}\left(\Omega_{1}-\Omega_{2}\right)
-\delta_{M,0}\Omega_{0}\right\}\delta_{G_\mu ,1}\ ,
\ee
where
\be
Q_\mu \equiv\frac{1}{\epsilon_{\rm v}-\epsilon_\mu}
\int \ dr\ r^2 \left\{\gvr\guiir\ +\fvr\fuiir\right\}\ .
\ee
Defining the following combinations, 
\be
\fupq&=&Q_\mu \fup\ ,\\
\gupq&=&Q_\mu \gup\ ,\quad  i=1,2
\ee
the isoscalar(vector) contributions to the spin polarized 
structure functions, eqs. (\ref{g10})--(\ref{gt1}), 
read
\be
\hspace{-1cm}
g^{I=0}_{1,\pm}(x,\mu^{2})&=&-N_C\frac{5M_N}{36\pi}
\int^\infty_{M_N|x_\mp|}p dp 
\nonumber \\* && \hspace{.1cm}
\times
\left\{\gvp\gip\frac{1-3\csii}{\sqrt{8}}-\gvp\giip\frac{1}{2}
\right.
\nonumber \\* && \hspace{.1cm}
\mp\gvp\fip\frac{\csi}{\sqrt{2}}
\mp\gvp\fiip\frac{\csi}{2}
\nonumber \\* && \hspace{.1cm}
\mp\fvp\gip\frac{\csi}{\sqrt{2}}
\mp\fvp\giip\frac{\csi}{2}
\nonumber \\* && \hspace{.1cm}
\left.
-\fvp\fip\frac{1+\csii}{\sqrt{8}}
+\fvp\fiip\frac{1-2\csii}{2}\right\}\ ,
\label{g1zro}
\ee
\be
\hspace{-1cm}
g^{I=1}_{1,\pm}(x,\mu^{2})&=&-N_C\frac{M_N}{36\pi}
\int^\infty_{M_N|x_\mp|}p dp 
\nonumber \\* && \hspace{.1cm}
\times\left\{\gvp^2\pm 2\gvp\fvp\csi
-\fvp^2\left(1-2\csii\right)\right\}\ ,
\label{g1one}
\ee
\be
\hspace{-1cm}
g^{I=0}_{T,\pm}(x,\mu^{2})&=&-N_C\frac{5M_N}{36\pi}
\int^\infty_{M_N|x_\mp|}p dp 
\nonumber \\* && \hspace{.1cm}
\times\left\{\gvp\gip\frac{3\csii-1}{4\sqrt{2}}-\gvp\giip\frac{1}{2}
\right.
\nonumber \\* && \hspace{.1cm} 
\left.
\mp\gvp\fip\frac{\csii-3}{4\sqrt{2}}
+\fvp\fiip\frac{\csii}{2}\right\}\ ,
\label{gtzro}
\ee
\be
\hspace{-1cm}
g^{I=1}_{T,\pm}(x,\mu^{2})&=&-N_C\frac{M_N}{36\pi}
\int^\infty_{M_N|x_\mp|}p dp 
\left\{\gvp^2 - \fvp^2\csii\right\}\ .
\label{gton}
\ee
which we evaluate numerically. Note that in case of the neutron the 
signs of the isovector pieces have to be reversed.
Note that the angle $\Theta_p^\pm$ is related to the integration
variable $p$ via
\be
{\rm cos}\Theta_p^\pm = \frac{1}{p}|M_N x \mp \epsilon_{\rm v}| .
\ee
In ref \cite{Di96} structure function $g_1$ was calculated omitting the 
cranking corrections. In the special case of the isovector component 
these corrections drop out and we formally confirm the result displayed
in eq (B.6) of ref \cite{Di96}.
\bigskip
\bigskip

\section{Evolution of 
${\overline{\lowercase{g}}}_{2}
\left(\lowercase{x},\mu^{2}\right)$}
\bigskip

In the $N_c\to \infty$ limit it has been shown\cite{Ali91}
that one can evolve the moments of 
${\overline{g}}_{2}\left(x,\mu^{2}\right)$,
\be
M_{j}
\left(Q^2\right)=\left(\frac{\alpha_{s}\left(Q^2\right)}{\alpha_{s}\left(\mu^2\right)}\right)^{\frac{\gamma_{j-1}}{b}}M_{j}\left(\mu^2\right)
\label{ali}
\ee
from the scale, $\mu^2$ to $Q^2$, where the anomalous dimensions are
\be
\gamma_{j-1}=2N_c\left(\psi (j)+\frac{1}{2j} + \gamma_{\rm E} -
\frac{1}{4}\right)\ ,
\ee
with $\psi(x)=\left(d/dx\right)\ {\rm log}\ \Gamma (x)$ and
$b=\left(11N_c -2n_f\right)/3$. $N_c$ and $n_f$ 
are the number of colors and flavors respectively.
Given the moments of ${\overline{g}}_{2}\left(x,\mu^{2}\right)$
\be
M_j\left(\mu^2\right)=\int^1_{0}\ dx\ x^{j-1}{\overline{g}}_{2}
\left(x,\mu^{2}\right)\ ,
\label{mt}
\ee
and expressing ${\overline{g}}_{2}\left(x,\mu^{2}\right)$ 
in terms of the ${\rm log}(x)$ and a power series in $x$,
\be
{\overline{g}}_{2}\left(x,\mu^{2}\right)=a_1\left(\mu^2\right)
{\rm log}(x)+\sum_{n=0}^{\infty} a_n\left(\mu^2\right) x^n
\ee
one can alternatively express the moments $M_j$ in terms of 
the coefficients $a_n$
\be
M_j\left(\mu^2\right)=A_{jn}a_n\left(\mu^2\right)\ .
\ee
We calculate the moments, $M_j\left(\mu^2\right)$ from (\ref{mt}) and evolve
them according to (\ref{ali}).
Finally, inverting the matrix $A_{jn}$ we obtain the
evolved coefficients, $a_n\left(Q^2\right)$ which in turn yields
${\overline{g}}_{2}\left(x,Q^{2}\right)$ (see figure 4).

\vfil\eject

\newpage

\begin{table}
\caption{
Sum rules calculated from eqs. (\ref{bcs}--\ref{gas}) as
functions of the constituent quark mass $m$ in the NJL chiral--soliton
model.}
\vspace{1cm}
\begin{tabular}{lddd} 
\hline
& $m$ ({\rm MeV})  & 400  & 450  \\ \hline
Burkardt--Cottingham: & $\Gamma_2^{p}$ & 0 & 0 \\ 
Bjorken: & $\Gamma_1^{p}-\Gamma_1^{n}=g_{A}/6$ & 0.121 & 0.118 \\ 
Ellis--Jaffe: & $\Gamma_1^{p}$  &  0.149 & 0.139 \\ 
ELT: & $\Gamma_{\rm ELT}$ & 1.38$\times 10^{-2}$ & 
7.65$\times 10^{-3}$\\ 
Axial Singlet Charge: & $\Gamma_1^{p}+\Gamma_1^{n}=g_A^0$ & 0.638 & 
0.579 \\
\end{tabular}
\end{table}

\newpage

\centerline{\bf\large Figure Captions}

\bigskip

\noindent
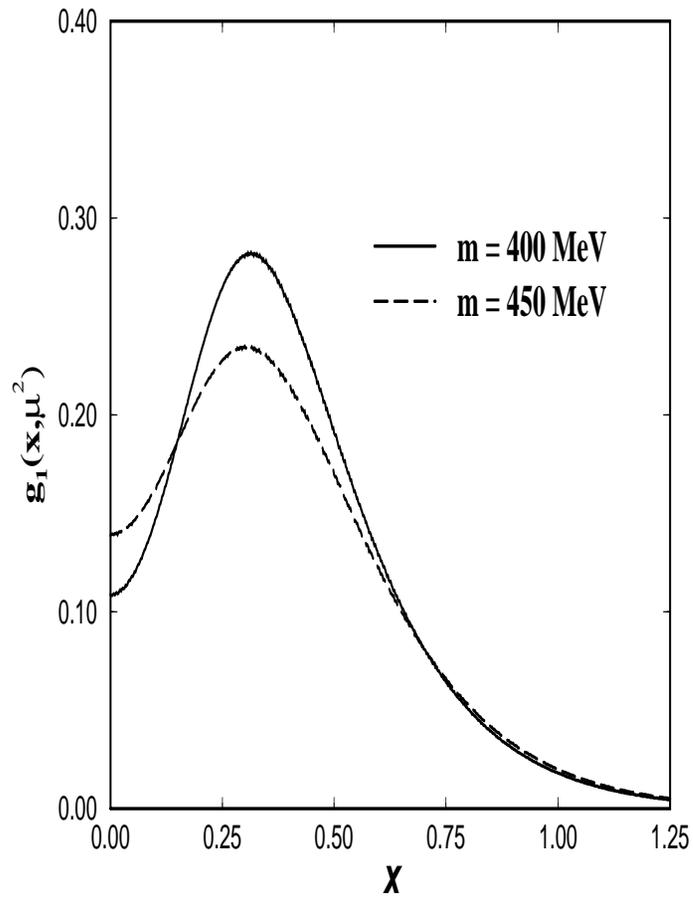
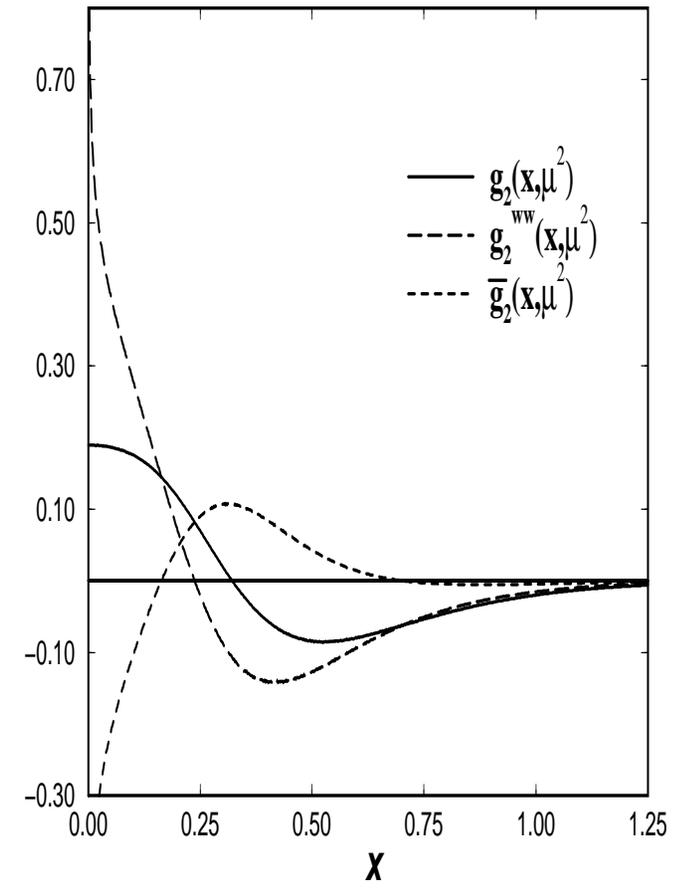
Figure 1: The valence quark approximation of the polarized proton
structure functions as a function of Bjorken--$x$.  Left panel:
$g_{1}^{p}\left(x ,\mu^{2}\right)$ for two constituent
quark masses $m$.  Right panel: $g_{2}^{p}\left(x
,\mu^{2}\right)$ (solid line), $g_{2}^{WW(p)}\left(x,\mu^{2}\right)$
(long--dashed  line) and twist three portion, ${\overline{g}}_{2}^{p}
\left(x,\mu^{2}\right)$ (dashed line). In this case we have used
$m=400{\rm MeV}$.

\bigskip

\noindent
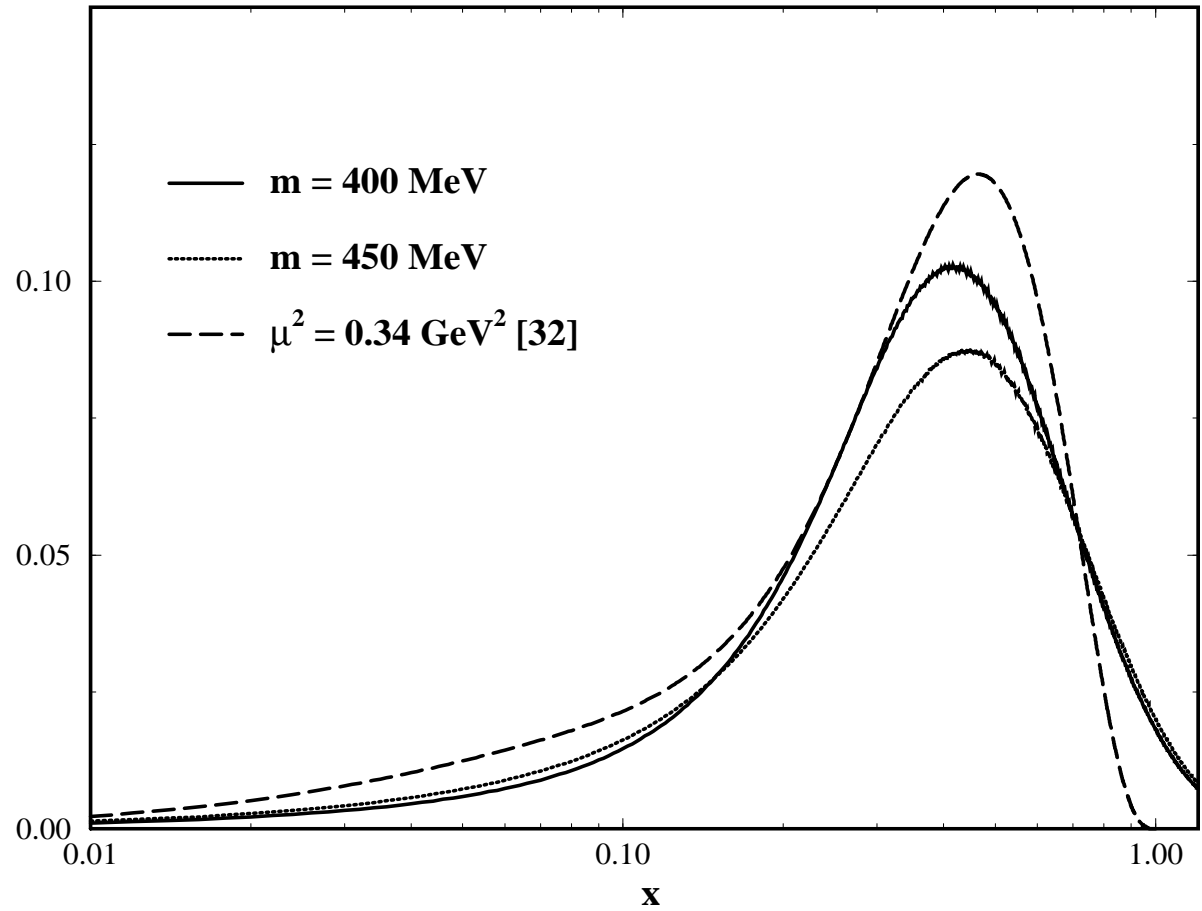
Figure 2: The valence quark approximation to the nucleon structure 
function $xg_1(x)$ in the NJL--soliton model compared to the 
low--renormalization point result of ref \protect\cite{Gl95}.

\bigskip

\noindent
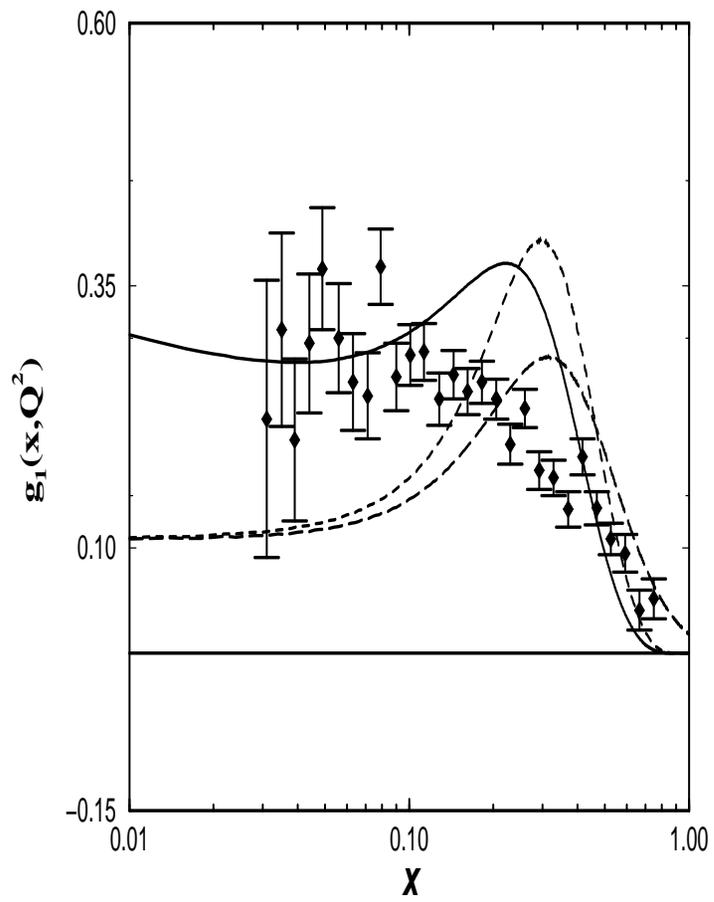
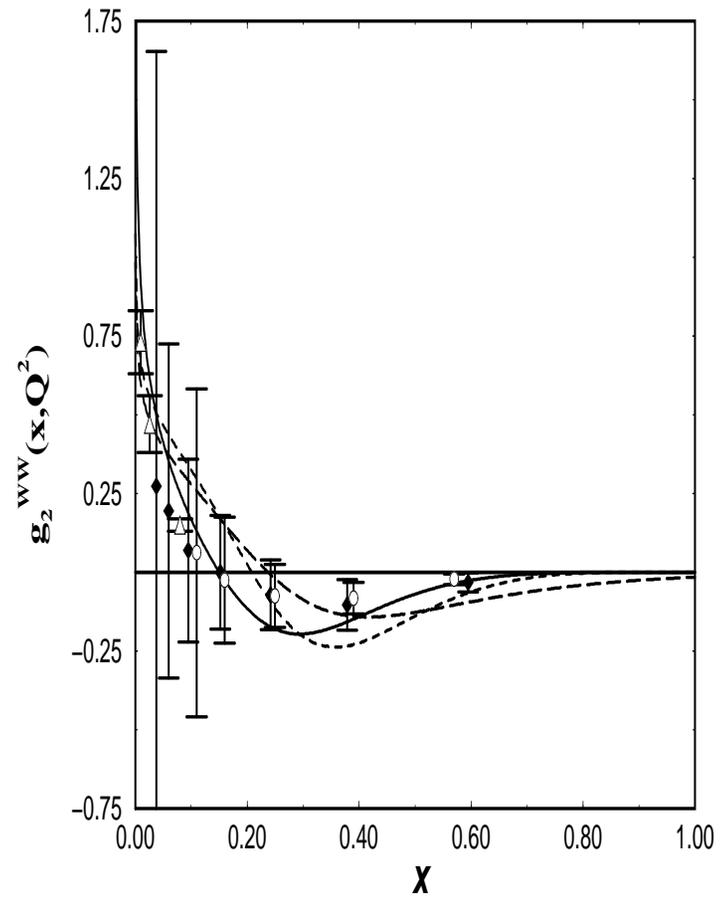
Figure 3: The projection and evolution of the spin--polarized structure
functions as a function of Bjorken--$x$.  For the constituent quark
mass we choose $m=400{\rm MeV}$.  Left panel:
$g_{1}^{p}\left({x},Q^2\right)$, unprojected (long--dashed line),
projected (dashed line) and evolved from $\mu =0.4{\rm GeV}^2$
to $Q^2=3.0{\rm GeV}^2$ (solid line). Data are
from \cite{slac95a}.
Right panel:  $g_{2}^{WW(p)}\left(x,Q^2\right)$,
unprojected (long--dashed line),
projected (dashed line) and evolved from $\mu =0.4{\rm GeV}^2$
to $Q^2=5.0{\rm GeV}^{2}$ (solid line).
Data are from \cite{slac96} and \cite{smc94b}, where
the diamonds, circles and triangles correspond to the
$4.5^\circ $ E143, $7.0^\circ $ E143 and SMC kinematics
respectively. Overlapping data have been shifted slightly in $x$.
The statistical error are displayed.

\bigskip

\noindent
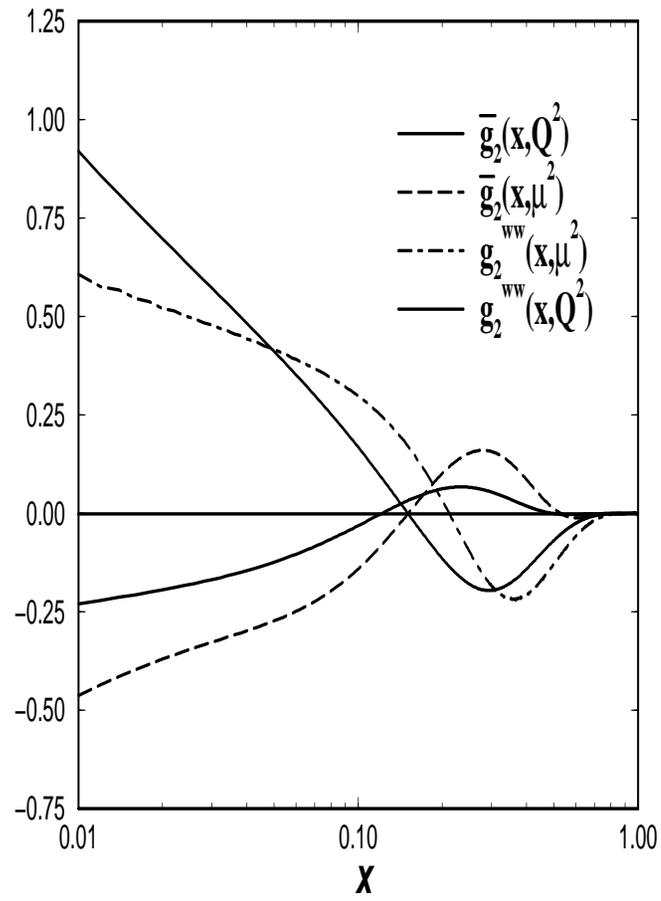
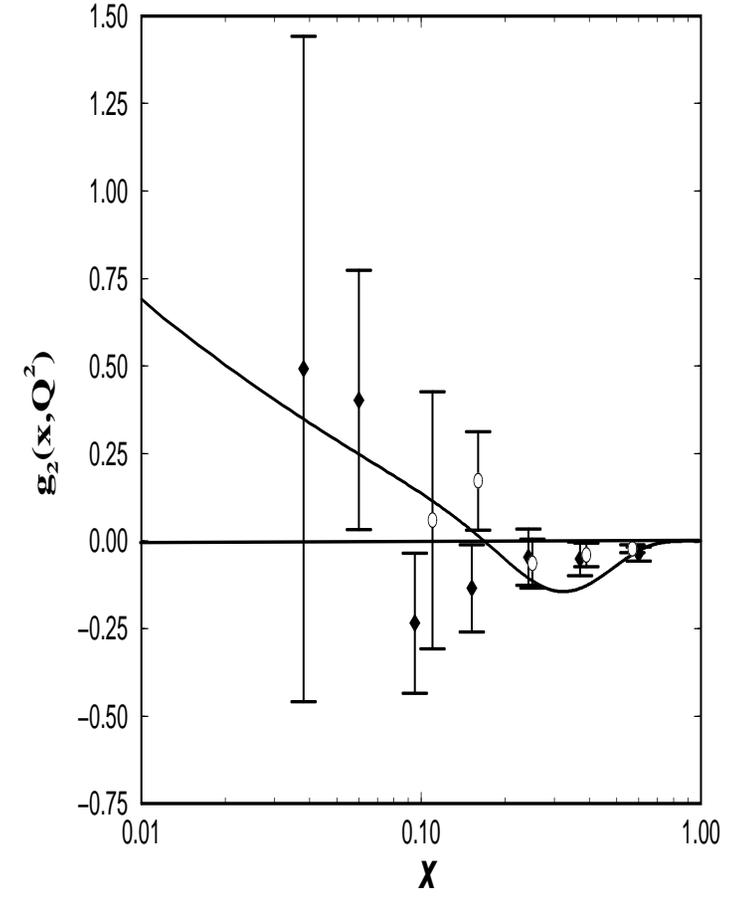
Figure 4: The evolution of
${\overline{g}}_{2}^{p}\left(x,Q^2\right)$ (projected)
from $\mu =0.4{\rm GeV}^2$ (long--dashed line) to $Q^2=5.0{\rm GeV}^2$
(solid line).  In addition we display the corresponding evolution
for $g_2^{WW(p)}\left(x,Q^2\right)$ (projected).
Right panel, $g_{2}^{p}\left(x,Q^2\right)=g_2^{WW(p)}\left(x,Q^2\right)+
{\overline{g}}_{2}^{p}\left(x,Q^2\right)$ evolved
from $\mu^{2}=0.4{\rm GeV}^2$ to $Q^2=5.0{\rm GeV}^2$.
Data and statistical errors for $g_2^{p}\left(x,Q^2\right)$
are displayed from \cite{slac96}, where
the diamonds and circles correspond to the
$4.5^\circ $ E143, $7.0^\circ $ E143 kinematics
respectively. Overlapping data have been slightly shifted in $x$.

\end{document}